\UseRawInputEncoding
\documentclass[twocolumn]{aastex631}
\usepackage{lineno}

\usepackage{newtxtext,newtxmath}

\usepackage{amsmath}	% Advanced maths commands
\usepackage{latexsym}
\usepackage{amssymb}	% Extra maths symbols
\usepackage{gensymb}
\usepackage{flafter}
\usepackage{appendix}
\usepackage{natbib}
\graphicspath{{./}{figures/}}
\newcommand\target{{4U~1630--47}}
\newcommand\hxmt{{\it Insight}-HXMT}
\newcommand\RXTE{{\it RXTE}}

\def\lax{\ifmmode{_<\atop^{\sim}}\else{${_<\atop^{\sim}}$}\fi}
\graphicspath{{./}{figures/}}

\received{January 1, 2018}
\revised{January 7, 2018}
\accepted{\today}
\submitjournal{ApJ}

\shorttitle{The mHz Quasi-Regular Modulation of \target}
\shortauthors{Zi-xu Yang et al.}
\begin{document}

\title{An \hxmt\ view of the mHz quasi-regular modulation phenomenon in the black hole X-ray binary \target}
%\title{The mHz Quasi-Regular Modulation phenomenon of the 4U 1630-47 observed by \hxmt\ during 2021 outburst}

\correspondingauthor{Zi-Xu Yang}
\email{yangzx@ihep.ac.cn}
\correspondingauthor{Liang, Zhang}
\email{zhangliang@ihep.ac.cn}
\correspondingauthor{Qingcui, Bu}
\email{bu@astro.uni-tuebingen.de}

\author{Zi-xu Yang}
\affil{Key Laboratory for Particle Astrophysics, Institute of High Energy Physics, Chinese Academy of Sciences, 19B Yuquan Road, Beijing 100049, China}
\affil{University of Chinese Academy of Sciences, Chinese Academy of Sciences, Beijing 100049, China}

\author{Liang Zhang}
\affil{Key Laboratory for Particle Astrophysics, Institute of High Energy Physics, Chinese Academy of Sciences, 19B Yuquan Road, Beijing 100049, China}

\author{Yue Huang}
\affil{Key Laboratory for Particle Astrophysics, Institute of High Energy Physics, Chinese Academy of Sciences, 19B Yuquan Road, Beijing 100049, China}

\author{Qingcui Bu}
\affil{Institut f\"ur Astronomie und Astrophysik, Kepler Center for Astro and Particle Physics, Eberhard Karls Universit\"at, Sand 1, D-72076 T\"ubingen, Germany」}

\author{Zhen Zhang}
\affil{Key Laboratory for Particle Astrophysics, Institute of High Energy Physics, Chinese Academy of Sciences, 19B Yuquan Road, Beijing 100049, China}

\author{He-Xin Liu}
\affil{Key Laboratory for Particle Astrophysics, Institute of High Energy Physics, Chinese Academy of Sciences, 19B Yuquan Road, Beijing 100049, China}
\affil{University of Chinese Academy of Sciences, Chinese Academy of Sciences, Beijing 100049, China}

\author{Wei Yu}
\affil{Key Laboratory for Particle Astrophysics, Institute of High Energy Physics, Chinese Academy of Sciences, 19B Yuquan Road, Beijing 100049, China}
\affil{University of Chinese Academy of Sciences, Chinese Academy of Sciences, Beijing 100049, China}

\author{Peng-Ju Wang}
\affil{Key Laboratory for Particle Astrophysics, Institute of High Energy Physics, Chinese Academy of Sciences, 19B Yuquan Road, Beijing 100049, China}
\affil{University of Chinese Academy of Sciences, Chinese Academy of Sciences, Beijing 100049, China}

\author{Q. C. Zhao}
\affil{Key Laboratory for Particle Astrophysics, Institute of High Energy Physics, Chinese Academy of Sciences, 19B Yuquan Road, Beijing 100049, China}
\affil{University of Chinese Academy of Sciences, Chinese Academy of Sciences, Beijing 100049, China}

\author{L. Tao}
\affil{Key Laboratory for Particle Astrophysics, Institute of High Energy Physics, Chinese Academy of Sciences, 19B Yuquan Road, Beijing 100049, China}

% ...



\begin{abstract}

Here we report the spectral-timing results of the black hole X-ray binary \target\ during its 2021 outburst using observations from the Hard X-ray Modulation Telescope (\emph{Insight}-HXMT). Type-C quasi-periodic oscillations (QPOs) in $\sim$1.6--4.2 Hz and quasi-regular modulation (QRM) near 60 mHz are detected during the outburst. The mHz QRM has a fractional rms of $\sim$10\%--16\% in the 8--35 keV energy band with a Q factor (frequency/width) of $\sim$2--4. Benefiting from the broad energy band of \hxmt, we study the energy dependence of the $\sim$60 mHz QRM in 1--100 keV for the first time. We find that the fractional rms of the mHz QRM increases with photon energy, while the time lags of the mHz QRM are soft and decrease with photon energy. Fast recurrence of the mHz QRM, in a timescale of less than one hour, has been observed during the outburst. During this period, the corresponding energy spectra moderately change when the source transitions from the QRM state to the non-QRM state. The QRM phenomena also shows a dependence with the accretion rate. We suggest that the QRM could be caused by an unknown accretion instability aroused from the corona.

\end{abstract}



\section{Introduction} \label{secintro}

 Black hole X-ray binary \target\ is a transient source characterized by its frequent outbursts \citep{2015MNRAS.447.3960C}. The first known outburst was recorded by Vela-5B in 1969 \citep{1986Ap&SS.126...89P}. After that, the source goes into outbursts frequently with a typical period of 600--700 days
 %there has been an not strictly regular outburst every ~600 days for 4U 1630-47 
\citep{1976ApJ...210L...9J,1986Ap&SS.126...89P,1995ApJ...452L.129P}. No dynamical mass measurement has been made for \target\ because of the difficulties in identifying an optical or IR counterpart up to now. By scaling the correlation of the photon index of the hard spectral component with low-frequency quasi-periodic oscillations (LFQPOs) and mass accretion rate, \citet{2014ApJ...789...57S} estimated that the black hole mass as $\sim$10~$\rm M_{\sun}$ and the inclination angle as $\leqslant 70$ deg. A potential distance of $4.7-11.5$ kpc was obtained by investigating the dust-scattering halo created by \target\ \citep{2018_ApJ_neardis}. 
\citet{Liu_2021_spin} suggested a moderately high spin of $a_{*} = 0.817\pm{0.014}$ (90\% confident level) by fitting the reflection spectra of \hxmt\ during its 2020 outburst. This value is lower than what had been reported by \citet{King_2014_wind}, in which they suggested an extreme high spin of $a_{*} = 0.985^{+0.005}_{-0.014}$ and an inclination angle of $i=64^{+2}_{-3}$ deg by fitting the {\it NuSTAR} spectra.

Fast variability observed in the X-ray band of black hole transients allows us to investigate the properties of the innermost region of accretion flow \citep{2006csxs.book...39V,2014SSRv..183...43B}. Meanwhile, the spectral-timing analysis allows us to understand the underlying radiation process. The Fourier transform serves as a powerful tool to study the fast X-ray variability \citep{2019NewAR..8501524I}. LFQPOs with frequencies of 0.1--30 Hz are common in black hole X-ray binaries. Based on the shape of the power density spectrum (PDS), LFQPOs are divided into three types, namely, type-A, B and C QPOs \citep{2005ApJ...629..403C}. 
The property of LFQPOs is highly correlated with spectral states. Type-C QPOs usually appear in the low-hard state and hard-intermediate state, whereas type-B and type-A QPOs appear in the soft-intermediate state. 
%
%Not like the common type-C QPO, the appearance of type-B/type-A QPO is rare event. 
%
Rapid transitions between different types of QPOs are sometimes observed during the hard-to-soft transition \citep{2008_MNRAS_Soleri,2011_MNRAS_Motta,2013ApJ...775...28S,2021_MNRAS_Zhangliang}. 
The physical origin of the LFQPOs is still under debate. It is suggested that a precessing hot inner flow/jet base can modulate the X-ray emission and produce the type-C QPO \citep{2007_A&ARv_Done,2009_MNRAS_Ingram,2010_MNRAS_Ingram,2011_MNRAS_Ingram,2016MNRAS.460.3284K,2018ApJ...866..122H,2021ApJ...919...92B,2021NatAs...5...94M,2022NatAs...6..577M}. This is currently the most promising model for type-C QPO.
No comprehensive models have been proposed for the type-B and type-A QPOs. \iffalse In addition to the LFQPOs, some other persistent BH XRBs exhibit exotic timing properties in high/soft state such as the mHz QPO in Cyg X-1 \citep{2021ApJ...919...46Y}, LMC X-1 \citep{1989PASJ...41..519E,2014MNRAS.445.4259A} and Cyg X-3 \citep{1985Natur.313..768V,2011MNRAS.416L..84K}.\fi

In addition to different types of QPOs, we have observed quasi-regular flares or dips with a relatively long period of $\sim$10--200 s in the light curves of a handful of black hole candidates, e.g., \target\ \citep{2001MNRAS.322..309T}, GRS 1915+015 \citep{1997ApJ...482..993M}, GRO J1655--40 \citep{1999ApJ...522..397R}, IGR J17091--3624 \citep{2011_ApJ_IGRJ17091,2013_IAUS_Zhangz,2014_A&A_Zhangzhen}, H 1743--322 \citep{2012ApJ...754L..23A}. These variability patterns show in the PDS as a broad peak at several tens of mHz. \citet{2001MNRAS.322..309T} named these signals as quasi-regular modulations (QRM).
%
%{\color{red}\textbf{The flux modulation, namely QRM, is characterized by a very particular type of light curve, which is significantly different from the type that characterizes LFQPOs. In general, the QRM can be directly identified by its regular flux modulation, whereas LFQPOs is usually observed in the frequency domain after performing Fourier transform.}} \iffalse However, they exhibit properties that differ from other QPO types, for instance, they occur only in a narrow range of X-ray luminosity \citep{2001A&A...372..138R}.\fi 
%
Taking GRS 1915+015 as an example, GRS 1915+015 is characterized by its complex X-ray variability which can be divided into 14 classes \citep{Belloni_2000_12,2002_MNRAS_2,2005_Hannika_new}. One of the most focused variability classes is the so-called "$\rho$" class (or "heartbeat" state). "Heartbeat" state is characterized by its regular flares which last 40--200 s \citep{2018_weng_GRS1915}. During the flare period, hardness ratio between different energy bands shows complex signatures \citep{2012_neilsen_radiation}. Based on a phase-resolved study, \citet{2012_neilsen_radiation} found that the temperature at the inner edge of the accretion disk increases with flux while the inner radius of the disk decreases during the flares. All these results support a radiation-pressure-driven evaporation or ejection event occurred in the inner accretion disk. At the inner region of the accretion disk where radiation pressure dominates, limit-cycle behaviour of accretion disk will cause periodic flux modulation due to the thermal viscous instability when the mass accretion rate increases to a certain level \citep{1974_ApJ_Lightman}. The instability is usually used to explain the accretion rate dependent flux modulation. Based on a systematic analysis of the "heartbeat" state in GRS 1915+105, \citet{2018_weng_GRS1915} found tight correlations among the recurrence time, the inner radius of the disk, and the luminosity of the non-thermal emission. They suggested that the change of the corona size can result in the observed correlations.

\citet{2000_ApJ_4U1630} and \citet{2001MNRAS.322..309T} studied the \RXTE\ observations of \target\ during its 1998 outburst. They found that during a certain plateau state ($\sim$1.4$\times10^{-8}~\rm erg~s^{-1}~cm^{-2}$, 3--20 keV energy range), the source exhibits quasi-regular oscillations with a period of $\sim$10--20 s with simultaneous complex type-C QPOs. \citet{2001MNRAS.322..309T} interpreted the type-C QPO by considering resonance oscillations of the shock, while the QRM is related to the time of the matter accumulation at the shock front (the shock stability timescale) . 

In 2021, \target\ experienced an outburst (peak$\sim$500 mCrab) after $\sim$600 days from its 2020 outburst. In this work, we report the detection of the mHz QRM phenomenon during this outburst, and present a detailed study of the mHz QRM from \target, in which we have extended its energy dependence study up to 60--100 keV for the first time. In addition, we compare the spectra of the periods with different types of PDS to unveil the spectral parameters that are responsible for the transition. \iffalse All the ObsIDs we selected are in the hard state (\citet{2006_McC_bhb} where the power-law contribution is less than 25 percent and (I am lost here...) > 80 percent in order to be classified as a soft and hard accretion rate, respectively.\fi In Section 2, we describe the observation and data reduction process. The data analysis and results are presented in Section 3. We discuss our results in Section 4.

\section{OBSERVATIONS AND DATA REDUCTION}

\hxmt\ is China's first X-ray astronomy satellite, launched on 2017 June 15 \citep{2020SCPMA..6349502Z}.
It carries three slat-collimated instruments: the High Energy X-ray telescope (HE: 20--250 keV, \citealt{2020SCPMA..6349503L}), the Medium Energy X-ray telescope (ME: 5--30 keV, \citealt{2020SCPMA..6349504C}), and the Low Energy X-ray telescope (LE: 1--15 keV, \citealt{2020SCPMA..6349505C}). Each telescope carries both large and small field-of-view (FoV) detectors in order to facilitate the background analyses. The small FoV detectors (LE: $1\degree.6 \times 6\degree$; ME: $1\degree \times 4\degree$; HE: $1\degree.1 \times 5\degree.7$) have a lower probability of source contamination, and thus are more suitable for pointing observation\footnote{\url{http://hxmten.ihep.ac.cn/AboutHxmt.jhtml}}.
\hxmt\ started  a high cadence monitoring of \target\ from 2021 September 18 to 28 (see Table~\ref{log} for the log of the \hxmt\ observation). The \hxmt\ observations stopped after 2021 September 28 due to the small solar aspect angle ($< 70 \degree$). 
%
\iffalse The LE (1--10 keV), ME (10--30 keV) and HE (30--150 keV) light curves of this outburst are shown in Figure~\ref{LEMEHE_lc}.\fi

\startlongtable

  \begin{deluxetable*}{lcccccccc}
   \tablecaption{\hxmt\ observations of \target\ in the 2021 outburst. \label{log}}
\tablehead{
\colhead{ObsID} & \colhead{Start Time} & \colhead{LE Rate} & ME Rate & HE Rate & \colhead{PDS\tablenotemark{a}} & \colhead{Centroid Frequency} & \colhead{Fractional rms\tablenotemark{b}} & \colhead{Q factor\tablenotemark{c}}\\
\colhead{} & \colhead{(MJD)} & \colhead{(cts $\rm s^{-1}$)} & \colhead{(cts $\rm s^{-1}$)} & \colhead{(cts $\rm s^{-1}$)} & \colhead{} & \colhead{(Hz)} & \colhead{(\%)} & \colhead{}
}
\startdata
  P040426300101 & 59475.647 & $5.2\pm{0.1}$ & $32.8\pm{0.2}$ & $74.5\pm{0.2}$ & QPO & $1.68^{+0.02}_{-0.03}$ & $20.63^{+1.58}_{-3.73}$ & $4.88^{+1.24}_{-1.94}$\\
  P040426300201 & 59476.488 & $7.6\pm{0.1}$ & $36.7\pm{0.1}$ & $93.8\pm{0.1}$ & QPO & $2.43^{+0.01}_{-0.01}$ & $18.20^{+0.74}_{-0.74}$ & $9.33^{+1.35}_{-1.14}$\\
  P040426300301 & 59477.647 & $10.3\pm{0.2}$ & $36.0\pm{0.2}$ & $92.9\pm{0.2}$ & QPO & $2.88^{+0.02}_{-0.02}$ & $16.82^{+1.70}_{-1.41}$ & $11.54^{+5.04}_{-3.32}$ \\
  P040426300401 & 59478.906 & $13.9\pm{0.1}$ & $39.9\pm{0.1}$ & $93.5\pm{0.2}$ & QPO & $3.55^{+0.04}_{-0.04}$ & $16.18^{+1.06}_{-1.06}$ & $6.83^{+1.55}_{-1.21}$\\
  P040426300501 & 59479.647 & $25.4\pm{0.3}$ & $64.7\pm{0.2}$ & $125.7\pm{0.3}$ & QPO & $4.28^{+0.16}_{-0.23}$ & $7.70^{+1.94}_{-2.05}$ & $6.48^{+6.29}_{-5.29}$\\
  P040426300601 & 59480.641 & $28.0\pm{0.3}$ & $61.6\pm{0.2}$ & $120.0\pm{0.3}$ & QPO & $4.17^{+0.05}_{-0.05}$ & $8.17^{+1.31}_{-1.20}$ & $16.68^{+10.49}_{-6.81}$\\
  P040426300701 & 59481.667 & $10.0\pm{0.2}$ & $34.7\pm{0.2}$ & $80.4\pm{0.3}$ & QPO & $2.35^{+0.05}_{-0.06}$ & $15.80^{+2.09}_{-2.43}$ & $9.02^{+3.46}_{-4.14}$\\
  P040426300801 & 59482.626 & $13.7\pm{0.1}$ & $40.2\pm{0.2}$ & $80.5\pm{0.2}$ & QPO & $3.32^{+0.07}_{-0.08}$ & $18.92^{+1.95}_{-1.78}$ & $3.68^{+1.33}_{-0.97}$\\
  P040426300901 & 59483.785 & $48.6\pm{0.3}$ & $112.5\pm{0.3}$ & $192.0\pm{0.9}$ & QRM & $0.062^{+0.001}_{-0.001}$ &$12.53^{+0.72}_{-0.72}$ & $2.96^{+0.80}_{-0.49}$\\
  P040426300902 & 59483.888 & $48.2\pm{0.3}$ & $106.9\pm{0.2}$ & $159.8\pm{0.2}$ & QRM & $0.064^{+0.001}_{-0.001}$ &$14.32^{+0.68}_{-0.68}$ & $3.51^{+0.71}_{-0.46}$\\
  P040426300903 & 59484.022 & $48.4\pm{0.2}$ & $109.4\pm{0.3}$ & $175.5\pm{0.4}$ & QRM & $0.060^{+0.002}_{-0.002}$ & $12.04^{+0.57}_{-0.57}$ & $1.90^{+0.55}_{-0.28}$\\
  P040426301001 & 59484.932 & $55.2\pm{0.3}$ & $114.3\pm{0.2}$ & $163.7\pm{0.3}$ & QRM & $0.053^{+0.001}_{-0.001}$ & $14.41^{+0.62}_{-0.62}$ & $2.04^{+0.47}_{-0.29}$\\
  P040426301002 & 59485.050 & $53.4\pm{0.4}$ & $115.5\pm{0.3}$ & $181.0\pm{1.4}$ & QRM & $0.059^{+0.001}_{-0.001}$ & $10.98^{+1.26}_{-1.26}$ & $8.36^{+4.46}_{-3.86}$\\
  P040426301003 & 59485.240 & $67.9\pm{0.3}$ & $131.1\pm{0.2}$ & $160.2\pm{0.2}$ & None/QRM & $0.055^{+0.001}_{-0.002}$ & $16.54^{+1.05}_{-1.05}$ & $3.40^{+0.90}_{-0.57}$\\
  P040426301004 & 59485.373 & $75.0\pm{0.3}$ & $149.8\pm{0.3}$ & $166.1\pm{0.2}$ & None & None & None & None\\
  P040426301005 & 59485.506 & $80.7\pm{0.3}$ & $154.3\pm{0.3}$ & $181.2\pm{0.3}$ & None & None & None & None\\                
  P040426301006 & 59485.639 & $80.1\pm{0.3}$ & $148.5\pm{0.4}$ & $165.8\pm{0.5}$ & None & None & None & None\\
  P040426301007 & 59485.771 & $78.4\pm{0.3}$ & $149.9\pm{0.3}$ & $172.4\pm{0.5}$ & None & None & None & None\\
  P040426301008 & 59485.909 & $79.1\pm{0.3}$ & $152.0\pm{0.2}$ & $167.3\pm{0.2}$ & None & None & None & None\\
\enddata

\tablenotetext{a}{The properties of the QPO and QRM are measured from the PDS of ME 8--35 keV.}
\tablenotemark{b}{The fractional rms is measured for the QPO/QRM component.}
\tablenotetext{c}{The quality factor Q is defined as centroid frequency/FWHM.}

\end{deluxetable*}

\iffalse
\begin{figure}
\centering
	\includegraphics[width=0.48\textwidth]{4U1630_lc_exp.pdf}
    \caption{\hxmt\ LE(1--10 keV), ME(10--30 keV), and HE(30--150 keV) light curves for 4U1630-47 in 2021 outburst.
    }   
    \label{LEMEHE_lc}
\end{figure}
\fi

\iffalse
\begin{figure}
\centering
	\includegraphics[width=0.48\textwidth]{MAXI_LC.pdf}
    \caption{Long-term light curves of \target\ in the 2--20 keV band obtained with MAXI/GSC.
    }
    \label{MAXI_lc}
\end{figure}
\fi

\begin{figure*}
\centering
	\includegraphics[width=\textwidth]{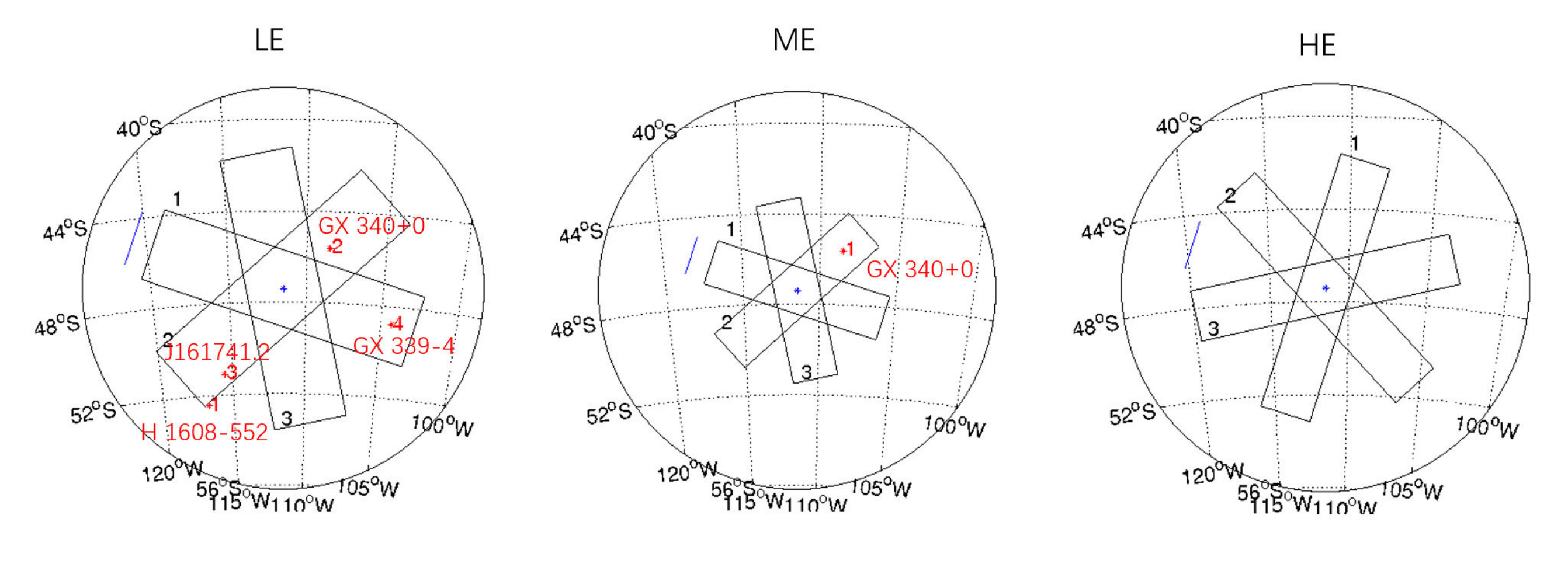}
    \caption{The small FoVs of LE (left), ME (middle) and HE (right). The locations and names of the contamination sources during our observations are marked with red. 
    %
    %GX 339-4 contributes considerable emission in LE detector, so the DetBox 1 of LE detector is discarded. GX 340+0 contributes non-negligible photon in LE and ME dector, so the DetBox 1,2 of LE detector and DetBox 2 of ME detector are discarded. 
    }
    \label{comtamination}
\end{figure*}

\begin{figure}
\centering
	\includegraphics[width=0.48\textwidth]{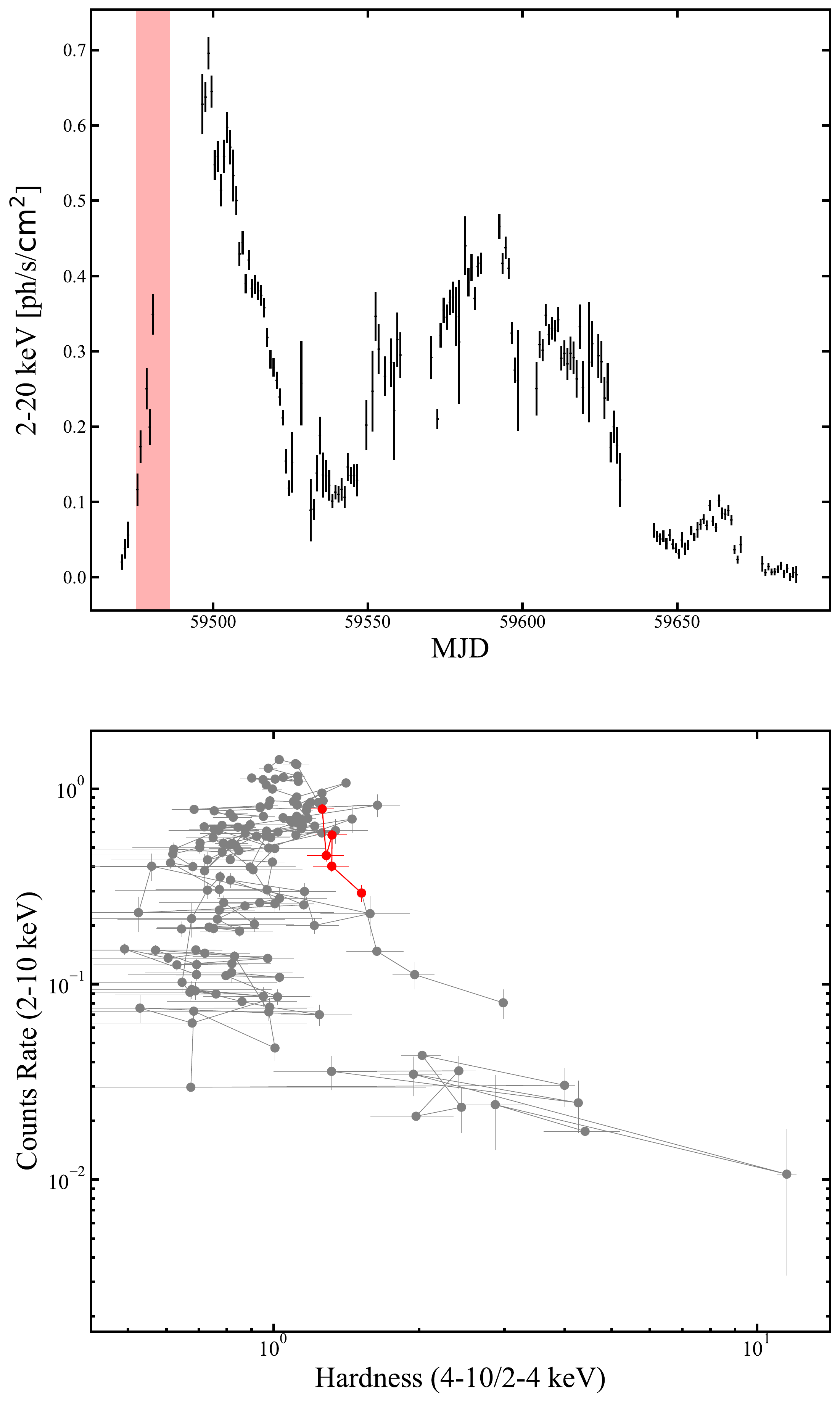}
    \caption{\textit{Upper panel}: The 2--20 keV light curve of \target\ in its 2021 outburst obtained with MAXI/GSC. The red region marks the time interval with \hxmt\ observations (MJD 59475--59486). \textit{Bottom panel}: the hardness-intensity diagram of \target\ in its 2021 outburst obtained with MAXI/GSC. The red points highlight the observations located in the red region of the upper panel.  
    }
    \label{MAXI}
\end{figure}

\begin{figure}
\centering
	\includegraphics[width=0.48\textwidth]{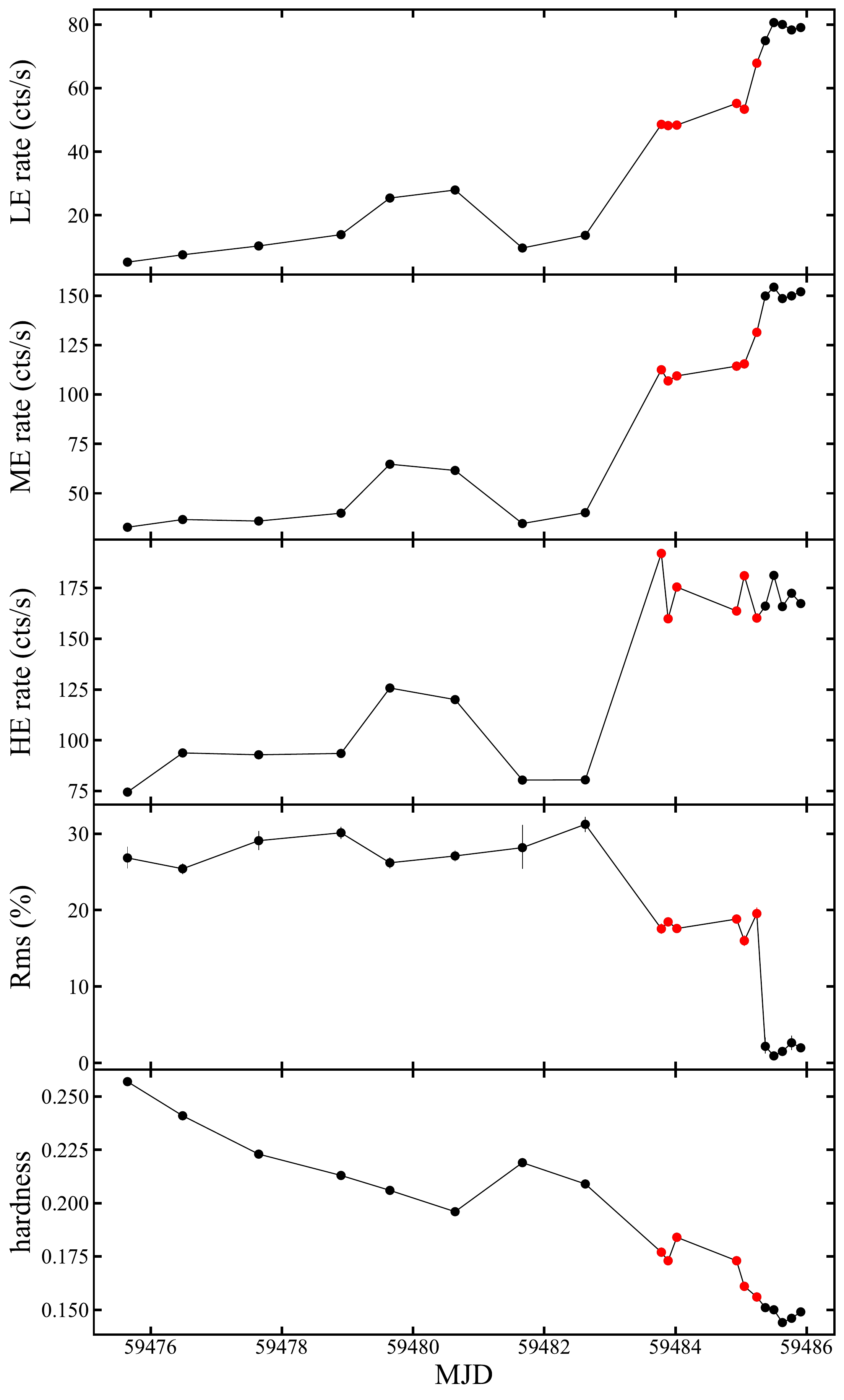}
    \caption{\hxmt\ LE (1--10 keV), ME (8--35 keV), and HE (30-150 keV) light curves, together with the 0.01--64 Hz fractional rms, calculated in the 8--35 keV (ME) band, and the hardness ratio defined as the ratio of the count rates between the ME 20--30 keV and ME 10--20 keV bands. The red points mark the observations with the mHz QRM.
    }
    \label{HXMT}
\end{figure}

\begin{figure*}
\centering
	\includegraphics[width=0.98\textwidth]{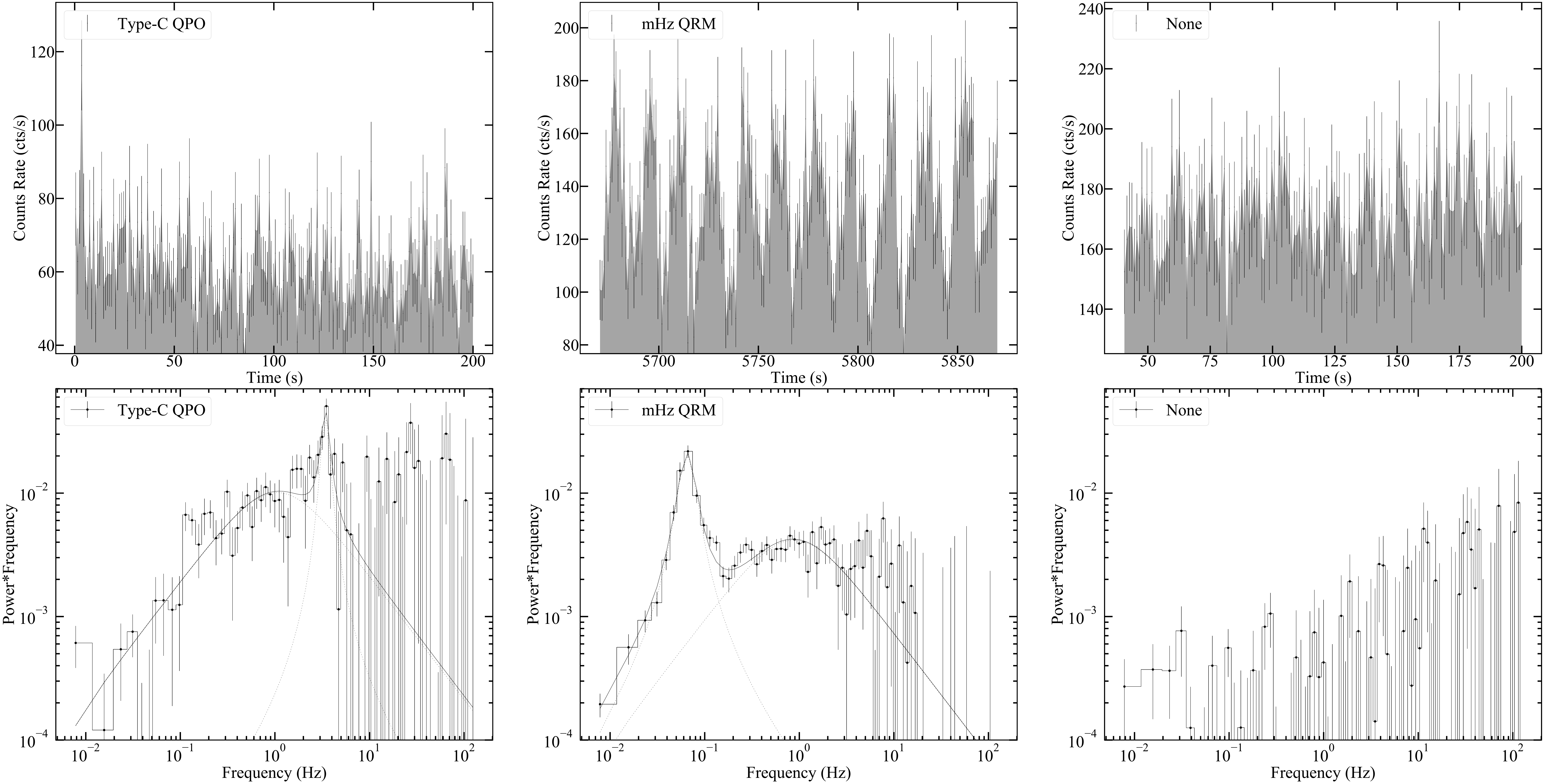}
    \caption{Representative light curves (ME 8--35 keV) and the corresponding power density spectra for different classes of X-ray variability observed in \target\ with \hxmt. \textit{Left panel:} the light curve and PDS with a typical type-C QPO. \textit{Middle panel:} the light curve and PDS with a mHz QRM. \textit{Right panel:} the light curve and PDS when no QPO/QRM is observed.
    %
    %Magenta points indicates the dates ObsID P0114661003 and P0114661004 were made (we use these two ObsIDs to investigate the energy dependence of broad-band noise). Gray dotted line marks the evolution phase which we choose to investigate the evolution trend with hardness in LHS.
    }
    \label{lc_pds}
\end{figure*}

\iffalse
\begin{figure}
\centering
	\includegraphics[width=0.48\textwidth]{MEpaper_QPO.pdf}
    \caption{A representative power density spectrum for ME (10--30 keV). The ObsID is P0404263008.
    %
    %Magenta points indicates the dates ObsID P0114661003 and P0114661004 were made (we use these two ObsIDs to investigate the energy dependence of broad-band noise). Gray dotted line marks the evolution phase which we choose to investigate the evolution trend with hardness in LHS.
    }
    \label{ME_paper_QPO}
\end{figure}

\begin{figure}
\centering
	\includegraphics[width=0.48\textwidth]{MEpaper_QRM.pdf}
    \caption{A representative power density spectrum for ME (10--30 keV). The ObsID is P0404263009.}
    \label{ME_paper_QRM}
\end{figure}
\fi

\begin{figure*}
\epsscale{1.0}
\centering
	\includegraphics[width=0.98\textwidth]{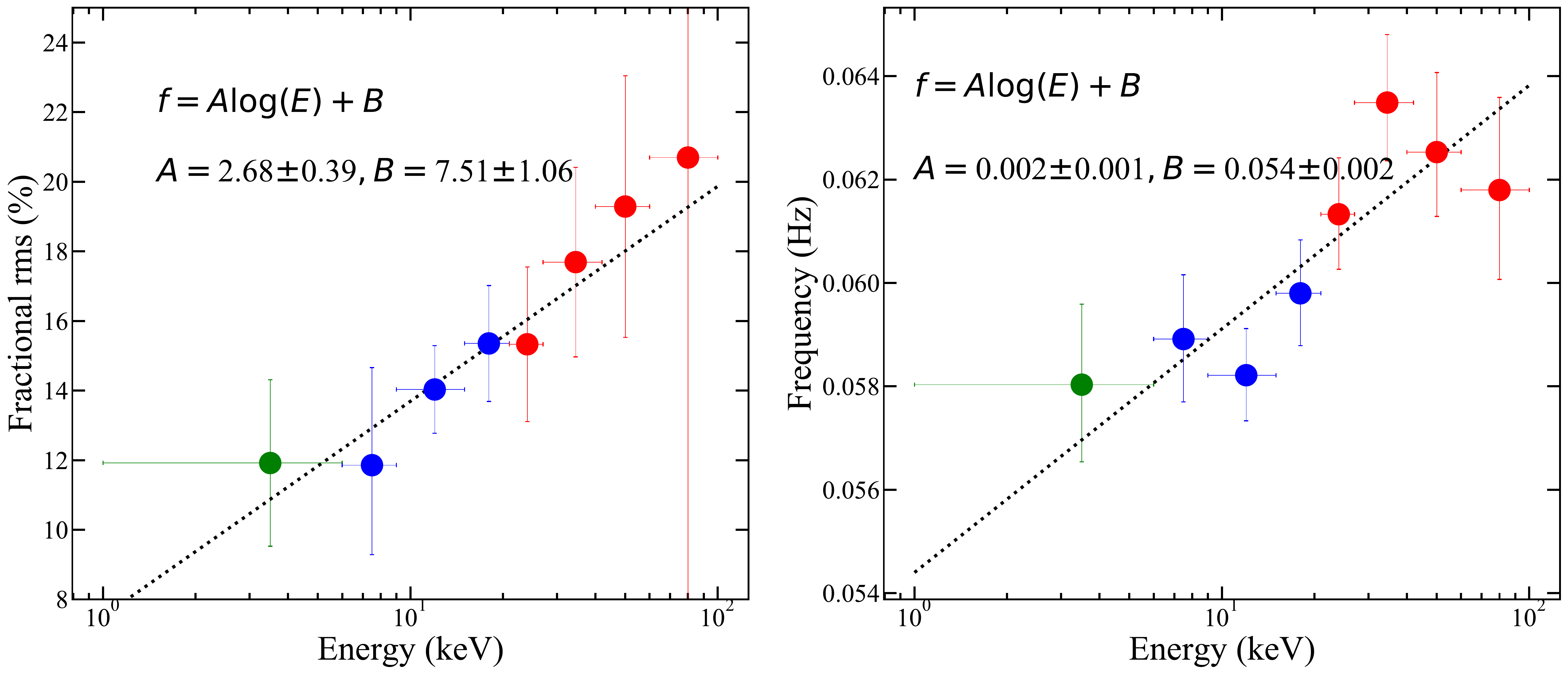}
    \caption{Fractional rms (left) and characteristic frequency (right) of the mHz QRM as a function of photon energy. The green, blue, red points represent the LE, ME and HE data, respectively. The dotted line represents the best-fitted line with a function of $f=A*{\rm log}(E)+B$. The fractional rms and characteristic frequency are measured from the PDS averaged from all the observations with the mHz QRM (ObsIDs P040426300901, P040426300902, P040426300903, P040426301001, P040426301002, P040426301003).}
    \label{E-rms-f}
\end{figure*}

\begin{figure*}
\epsscale{1.0}
\centering
	\includegraphics[width=0.98\textwidth]{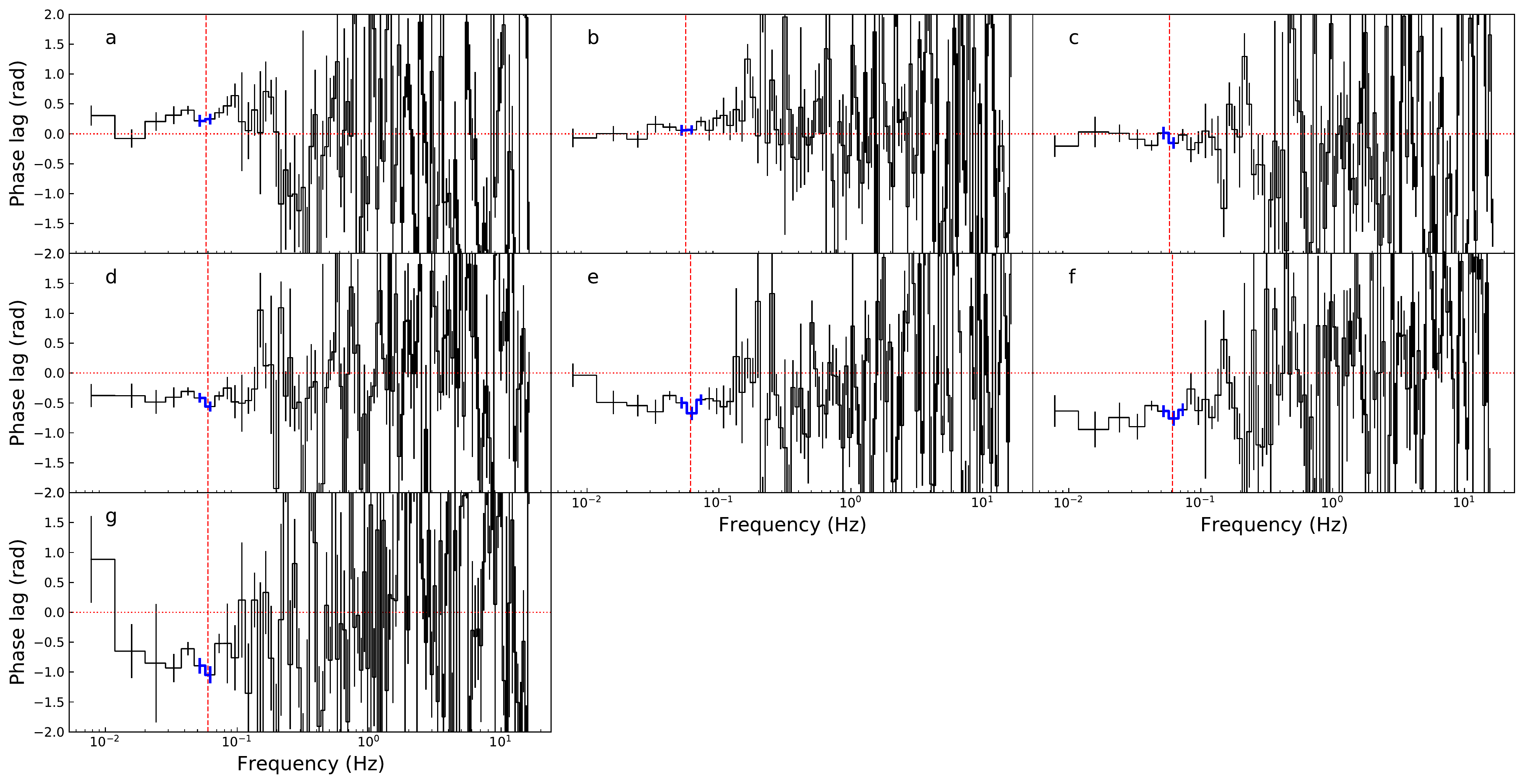}
    \caption{Frequency-dependent phase-lag spectra in different energy bands for the observations with the mHz QRM. The reference energy band is LE 1--6 keV. The chosen energy bands are 6--9 keV (a), 9--15 keV (b), 15--21 keV (c), 21--27 keV (d), 27--40 keV (e), 40--60 keV (f), and 60--100 keV (g). The red vertical line marks the centroid frequency of the mHz QRM. The blue points mark the frequency range ($\nu_{0}-{\rm FWHM}/2, \nu_{0}+{\rm FWHM}/2$) over which the phase-lags of the mHz QRM are averaged. Same as Figure~\ref{E-rms-f}, the data we used are from all the observations with the mHz QRM.}
    \label{frequency_lag}
\end{figure*}

\begin{figure}
\epsscale{1.0}
\centering
	\includegraphics[width=0.49\textwidth]{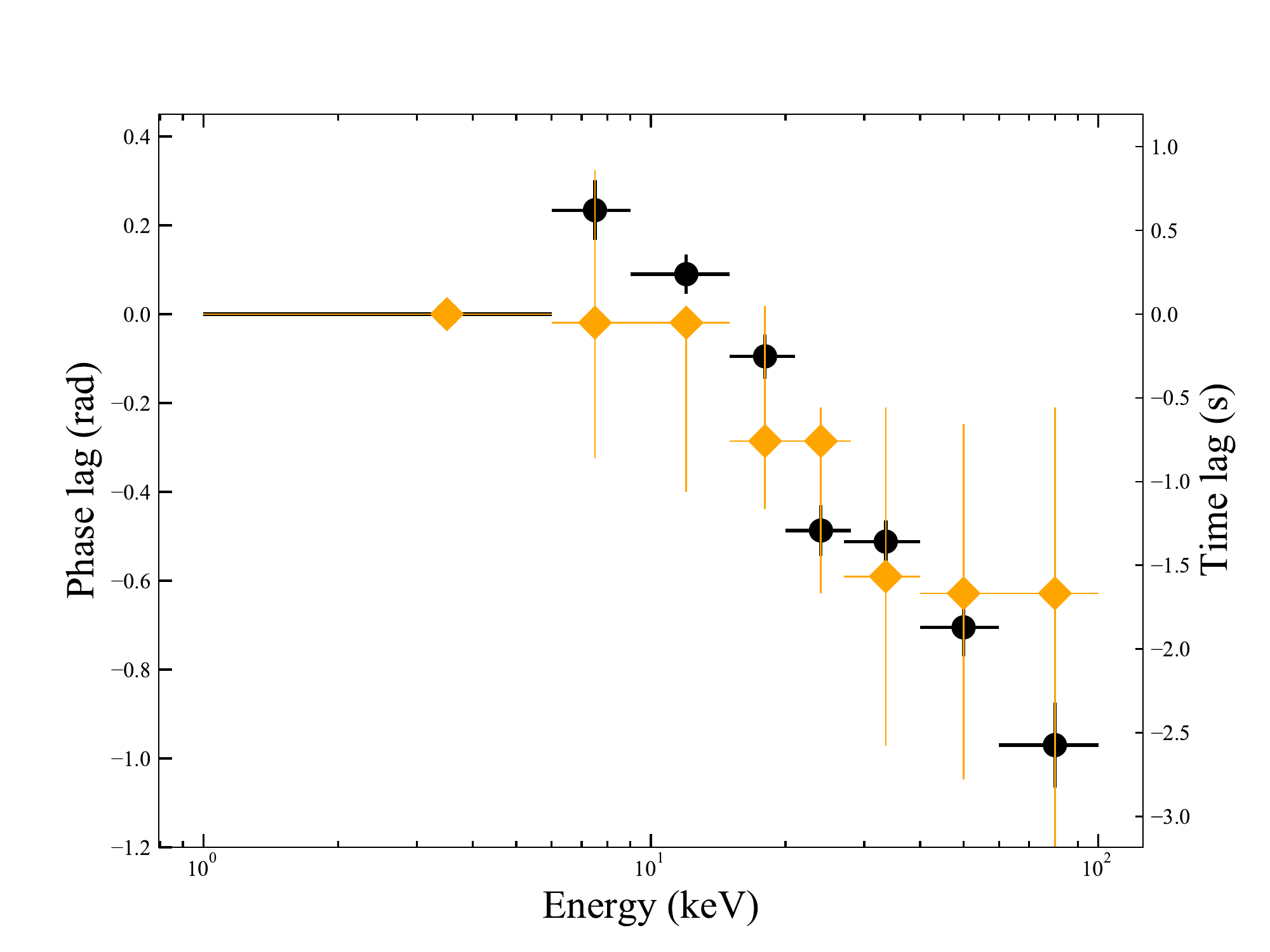}
    \caption{Phase/time lags as a function of photon energy. The reference energy band is LE 1--6 keV. The chosen energy bands are 6--9 keV, 9--15 keV, 15--21 keV, 21-27 keV, 27--40 keV, 40--60 keV, and 60--100 keV. The black points  are the phase/time lags of the mHz QRM measured from the frequency-dependent phase-lag spectra by averaging the lags in the QRM frequency range ($\nu_{0} \pm {\rm FWHM}/2$, blue points in Figure \ref{frequency_lag}). 
    The orange points are the time lags measured from cross-correlation in time domain of two different energy band light curves. \iffalse The phase lags are measured following the method described in \citet{2021NatAs...5...94M}. The black, blue and red points are the ``original'' phase lags of the mHz QRM, the lags of the broadband noise continuum, and the ``intrinsic'' lags of the mHz QRM, respectively. See the text for details.\fi}
    \label{timelag_evolution}
\end{figure}

\begin{figure*}
\epsscale{1.0}
\centering
	\includegraphics[width=0.98\textwidth]{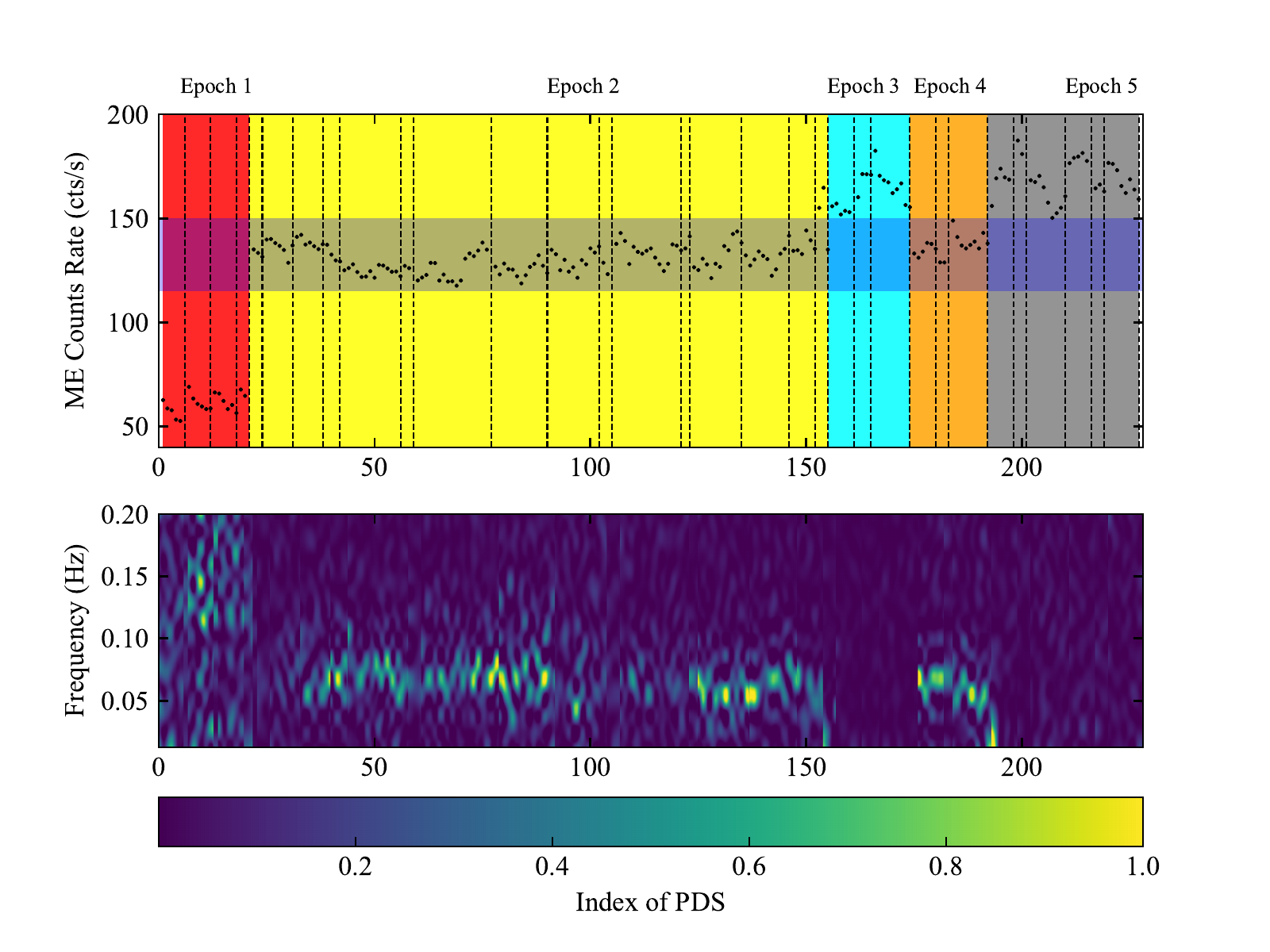}
    \caption{ME 8--35 keV light curve and the corresponding dynamical PDS for a particular period that shows fast transitions among different kinds of PDS. The bin size of the light curve is 80 s. In the top panel, the dashed lines mark the time gaps between orbits. In the bottom panel, the $x$-axis represents the index of each PDS. 
    We divide this period into five epochs: epoch 1 ( P040426300801), the period with a type-C QPO in the PDS; epoch 2 (P040426300901, P040426301001, P040426301002), the period shows a mHz QRM; epoch3 (the first part of P040426301003), the period when mHz QRM disappears and the PDS is dominated by Possion noise; epoch 4 (the last part of P040426301003), the period when the mHz QRM reappears, and epoch 5 (ObsID P040426301004) where QRM disappears again. }
    % The data we used include ObsID P0404263008 (red region), P0404263009 (yellow region), P040426301001 (yellow region), P040426301002 (yellow region), P040426301003 (blue region and orange region), P040426301004 (gray region). Red dashed region marks epoch 1 with the type-C QPO. Yellow dashed region marks epoch 2 with the mHz QRM. Cyan dashed region marks epoch 3 without the mHz QRM temporarily. Orange dashed region marks epoch 4 with the mHz QRM again. Gray dashed region marks epoch 5 without the mHz QRM and any QPO signal.}
    \label{dyps}
\end{figure*}

\iffalse
\begin{figure}
\epsscale{1.0}
\centering
	\includegraphics[width=0.48\textwidth]{E-width.pdf}
    \caption{Full width at half maximum as a function of photon energy for mHz QRM component.}
    \label{E-width}
\end{figure}
\fi

\iffalse
\begin{figure}
\epsscale{1.0}
\centering
	\includegraphics[width=0.48\textwidth]{E-lag.pdf}
    \caption{phase-lag at QRM frequency as a function of photon energy.}
    \label{E-lag}
\end{figure}
\fi

The data are extracted from all three instruments using the \hxmt\ Data Analysis software (HXMTDAS) v2.05 \footnote{The data analysis software is available from \url{http://hxmten.ihep.ac.cn/software.jhtml}.}, and filtered with the following criteria:  
(1) pointing offset angle less than $0.04\degree$;  
(2) Earth elevation angle larger than $10\degree$;  
(3) the value of the geomagnetic cutoff rigidity larger than 8 GV;  
(4) at least 300 s before and after the South Atlantic Anomaly passage.
To avoid possible contamination from the bright Earth and nearby sources, we only use data from the small field of view (FoV) \citep{2018ApJ...864L..30C}. The energy bands adopted for spectral analysis are 2--8 keV for LE, 8--28 keV for ME, and 28--100 keV for HE.

% Meanwhile, in order to constrain the interstellar medium grain absorption parameter, we also utilize one observation carried out by NICER, see Table~\ref{log}. The NICER observation correspond to the \hxmt\ observation with the mHz QRM simultaneously. After the event files are downloaded, the light curve and spectrum are generated with the multipurpose tool XSELECT v2.4k\footnote{https://heasarc.gsfc.nasa.gov/docs/software/lheasoft/ftools/xselect/}.  

\subsection{Background}
The background estimations of LE, ME, and HE are performed using the stand-alone Python scripts LEBKGMAP, MEBKGMAP, and HEBKGMAP \citep{2020JHEAp..27...14L,2020JHEAp..27...24L,2020JHEAp..27...44G}. \target\ locates in a crowded region near the Galactic plane. As a result, it may be contaminated by some other bright X-ray sources  \citep{1997MNRAS.291...81K}.
In Figure~\ref{comtamination}, we show the small FoVs of LE, ME, and HE, respectively. Each instrument consists of three Detection Boxes (DetBox No. 1, 2, and 3).
As can be seen, relatively bright contaminating sources in the small FoVs of \target\ are GX 340+0, GX 339--4, H1608--552 and J161741.2--510455. H1608--552 and J161741.2--510455 are transient sources that were in a quiescence state during the outburst of \target, thus their contribution can be ignored. 
GX 340+0 is a persistent Z source that shows fast X-ray variability. It constantly appears in the No.2 DetBox of the LE detector and No.2 DetBox of the ME detector. Following \citet{2021ApJ...919...92B}, We create light curves and spectra from each DetBox and find that the contribution of the X-ray flux from GX 340+0 is larger than 5\% for the LE and ME detectors. 
In addition, GX 339--4 was at the end stage of the outburst in 2021. It constantly appears in the No.1 DetBox of the LE detector, and contributes more than 50\% extra counts to the LE No.1 DetBox detector. Therefore, we do not use the data from the No.1 and 2 DexBox of the LE detector and the No.2 DetBox of the ME detector for the following timing and spectral analysis.

\section{ANALYSIS AND RESULTS} \label{sec:floats}

\subsection{Fundamental diagrams}
% First, we plot the diagrams commonly used for the study of black hole X-ray transients in Figure~\ref{MAXI} and Figure~\ref{HXMT} taken by MAXI/GSC  \citep{2009PASJ...61..999M} and \hxmt.

In the upper panel of Figure~\ref{MAXI}, we show the 2--20 keV light curve of \target\ from its 2021 outburst obtained with MAXI/GSC. Three peaks can be observed from the light curve. The \hxmt\ observations were performed during the initial rising phase of the first main peak.
In the lower panel of Figure~\ref{MAXI}, we show the hardness-intensity diagram (HID) of this outburst. Based on their locations in the HID, we can identify that our \hxmt\ observations are made during the intermediate state. In the 2021 outburst although we still lack bright hard states as previous outbursts \citep{2005PASJ...57..629A,2014ApJ...791...70T,2015MNRAS.450.3840C}, \hxmt\ accumulates intensive observations in hard-intermediate/steep power-law state which are rare during \RXTE\ epochs \citep{2005ApJ...630..413T,2021ApJ...909..146C}.
%
% As bottom panel shows, the 2021 outburst shows a fast rise-exponential decay (FRED) profile which is typical for black hole X-ray transients but along with two flux peaks \citep{1997ApJ...491..312C}.

%In the 2021 outburst, \hxmt\ caught the source in the hard-intermediate state/steep power-law state, which have been rarely observed in RXTE epochs \citep{2005ApJ...630..413T,2021ApJ...909..146C}. Most of the previous outbursts caught the source in the bright hard states \citep{2005PASJ...57..629A,2014ApJ...791...70T,2015MNRAS.450.3840C}.
%

In Figure~\ref{HXMT}, we show the \hxmt\ LE (1--10 keV), ME (8--35 keV), and HE (30--150 keV) light curves, together with the 0.01--64 Hz fractional rms, calculated in the ME (8--35 keV) band, and the hardness ratio, defined as the ratio of the count rates between the ME 20--30 keV and ME 10--20 keV bands, evolution during the outbursts. The LE/ME/HE count rates slowly increase from initial value, jump to a plateau around MJD 59483, and then remain stable at this flux level for about 3 days. Afterwards, the count rates rapidly increase to a higher flux level, accompanied by a decrease in hardness ratio. Meanwhile, at the beginning of the observation, the total fractional rms remains above 20\%, whereas after MJD 59483, the rms drops sharply to between 20\% and 10\%. After MJD 59485, the PDS shows a very low rms amplitude (<5\%).

\subsection{Timing analysis}

To study the fast X-ray variability, we create an averaged power spectrum for each of the \hxmt\ observations. Based on the characteristics of the light curves and PDS, we divide the X-ray variability into three different types. In Figure~\ref{lc_pds}, we show the representative light curves and their corresponding PDS for the three types, respectively. 
During MJD 59475.6--59482.6, the PDS (the left panels of Figure~\ref{lc_pds}) is characterized by a relatively narrow QPO centered at 1.6--4.2 Hz with a band-limited noise component at low frequencies.
During MJD 59483.8--59485.2, the PDS (the middle panels of Figure~\ref{lc_pds}) is characterized by a relatively broad peak ($Q \sim 2-4$) near 60 mHz accompanied with a broadband noise component above its frequency. The corresponding light curves show regular flares of $\sim$16 s, which are clearly different from the light curves of the first type. We further refer this regular modulated signal as mHz QRM to distinguish it from the first type. In addition, the mHz QRM shows a transient behaviour near MJD 59485: the mHz QRM disappears in the first two orbits of this observation (ObsID P040426301003) and appears again in the last two orbits.
After MJD 59485.4, the PDS is dominated by Possion noise without QPO/QRM.

For further studies, we create PDS with a 128 s long interval and 1/256 s time resolution. The PDS is applied to Miyamoto normalization \citep{1991ApJ...383..784M}.
We then fit the PDS with a multiple-Lorentzian model \citep{2002ApJ...572..392B} (see Figure 4). 
From the fittings, we obtain the fractional rms and characteristic frequency of the QPO/QRM. The fractional rms is background corrected according to the formular: $rms = \sqrt{P}\times(S+B)/S$ \citep{1990A&A...227L..33B,2015ApJ...799....2B}. The maximum power is observed at the characteristic frequency $\nu_{\rm max}=\sqrt{\nu_0^2+(\sigma/2)^2}$, where $\nu_0$ is the centroid frequency and $\sigma$ is the FWHM of the Lorentzian function \citep{1997A&A...322..857B}. The values of the frequency and rms of the QPO/QRM are listed in Table~\ref{log}.
%By visual inspection, we can easily find that compared with LE and ME PDS, there is obvious shape change at higher frequency (exceed QPO frequency) in HE PDS while at low frequency no obvious spectral shape change was found. In fact, we can also see slight difference at 1--10 Hz frequency between LE and ME PDS. This also means the broad-band noise component has energy dependence notably at HE band. 
%
%not only based on experience but also fluctuation propagation model \citep{2000MNRAS.318..361N,2002ApJ...572..392B}.
%

Two Lorentzian components are required to fit the PDS of the 1.6--4.2 Hz QPO: a narrow peak and a low-frequency broadband noise component. The frequency of the QPO is anti-correlated with spectral hardness and QPO fractional rms. The $Q$ factor of this QPO is in the range of $\sim$4--10, with a fractional rms amplitude of $\sim$8--21$\%$. The broadband noise component is relatively strong with a typical rms of 20$\%$. Based on these characteristics, we identify this QPO as the type-C QPO detected in black hole X-ray binaries.

The PDS of the mHz QRM can be well fitted with two Lorentzian functions. However, their properties are quite different from those of the 1.6--4.2 Hz QPOs. The frequency of the peak is around 0.06 Hz, and stays stable among different observations. The width of the peak is broad, with a $Q$ factor of $\sim$ 2--4. The fractional rms amplitude of the mHz peak is $\sim$11-16$\%$. The broadband noise component above the 0.06 Hz peak has a relatively lower rms of $\sim$10$\%$. 
From the view of the light curves, the mHz flux modulation is characterized by a fast rise following by a fast decay. The rise and decay times of the flare are almost the same ($\sim$8 s). This is different from the characteristics of the "heartbeat" light curves, which typically show a slow rise and fast decay profile \citep{2018_weng_GRS1915,2021A&A...650A.122M}.

We further study the energy-dependent properties of the mHz QRM. Since the properties of the mHz QRM are similar among observations, we combine all the observations with the mHz QRM detection and compute averaged PDS for different energy bands. The fractional rms and the characteristic frequency of the mHz QRM as a function of photon energy are shown in Figure~\ref{E-rms-f}. From the left panel of Figure~\ref{E-rms-f}, we can see a positive correlation between the fractional rms of the mHz QRM and photon energy in the 1--100 keV energy band. The value of the rms increases from $\sim$12\% to $\sim$20\%. In order to check the significance of the increase, we fit this relation with a function of $f=A*{\rm log}(E)+B$. The best-fitting slope is $A=2.68\pm0.39$, which is more than 6$\sigma$ from zero. 
%This indicates that the increasing trend in fractional rms with energy is statistically significant.
%
%Such a significant positive correlation has never been discovered in any other black hole transients \citep{2004ApJ...615..416R,2012ApJ...754L..23A,2013MNRAS.434...59Y,2013MNRAS.428.1704L,2013MNRAS.433..412L,2004A&A...426..587C,2016ApJ...833...27Y,2018ApJ...866..122H}. 
%
There is also a slightly increasing trend in characteristic frequency (the right panel of Figure~\ref{E-rms-f}). However, the change in characteristic frequency is relatively small, from $0.058^{+0.002}_{-0.001}$ Hz to $0.064^{+0.001}_{-0.001}$ Hz. We fit this relation with the function $f=A*{\rm log}(E)+B$. The best-fitting slope $A=0.002\pm0.001$ is consistent with being zero within $\sim$2$\sigma$ level, suggesting that there is no significant change in characteristic frequency. Meanwhile, it should be noted that the mHz QRM peak is broad, and its rms or frequency could be affected by the variation of the underlying broad band noise component. In particular, a strong energy dependence of the broad band noise component has been found in previous studies \citep{2022ApJ...932....7Y,2022arXiv220108588F}. 

Phase/time lag is a commonly used tool to study the X-ray variability and can help us understand the geometry of the emission area. In Figure~\ref{frequency_lag}, we show the  frequency-dependent phase-lag spectra for different energy bands, relative to the same reference band 1--6 keV. From this figure, we do not see clear features around the mHz QRM frequency range. The phase lags below 0.1 Hz remain more or less constant. This is consistent with the fact that the PDS below 0.1 Hz is completely dominated by the Lorentzian component of the QRM. This is in sharp contrast with that type-C QPOs, which are strongly interfered by the underlying broadband noise \citep{2021NatAs...5...94M}. Above 0.1 Hz, the phase lags can not be constrained well.  
%
\iffalse To calculate the ``intrinsic'' phase lag of the mHz QRM, we use the method described in \citet{2021NatAs...5...94M}.\fi 
%
We calculate the phase lags of the mHz QRM by averaging the lags over the mHz QRM frequency range $\nu_{0} \pm \rm FWHM/2$, as we commonly do for QPOs \citep{1997ApJ...474L..43V,2010ApJ...710..836Q,2017ApJ...845..143Z}. The results are shown in Figure~\ref{timelag_evolution} with black points. Below 10 keV, we find a slight hard lag or zero lag.  Above 10 keV, the lags are soft and decrease with energy up to 100 keV. 
Since the light curve is mainly dominated by the flare (QRM), we also calculate the time lags from the cross-correlation in time domain (see orange points in Figure~\ref{timelag_evolution}). Although the errors of the time lags measured from the cross-correlation are very large, we can see that the trend of the black and orange points is consistent.

\subsection{Spectral changes during the transitions between different kinds of PDS}

During the period between MJD 58482--58485, we observe fast transitions among different types of PDS, as shown in Figure~\ref{lc_pds}. In Figure~\ref{dyps}, we show the ME 8.0--35.0 keV light curve with 80-s binning and the corresponding dynamical PDS within this period. The dashed lines in the top panel mark the time gaps between orbits. The $x$-axis label is the index of each 80-s PDS. 
Based on the characteristic of the PDS and the flux level, we divide this period into 5 epochs. During epoch 1, significant type-C QPOs with a strong band-limited noise at low frequencies are observed in the PDS. The type-C QPO disappears in epoch 2 and is replaced by the mHz QRM. The mHz QRM disappears in epoch 3, re-appears in epoch 4, and then disappears in epoch 5.
%(epoch 1: red region; epoch 2: yellow region: epoch 3: cyan region; epoch 4: orange region; epoch 5: gray region). The red, yellow, cyan, orange and gray region respectively represent the epoch with the type-C QPO, mHz QRM, without the mHz QRM temporarily, with the mHz QRM again, and the epoch without the QPO and mHz QRM. 
It is clear that the mHz QRM is only observed within a narrow flux range (ME 8--35 keV count rates $\sim120-150 \ \rm cts \ s^{-1}$). The behaviour of the regular flux modulation disappears outside of this range.
%correspondingly average counts rate $\sim1.37\times 10^{-8} \rm erg \ \rm s^{-1} \ \rm cm^{-2}$, 3--20 keV energy range), the mHz QRM appears, namely, in epoch 2 and epoch 4. 

To study the spectral changes during the transitions among the epochs, we extract background-subtracted spectra for the 5 epochs, separately. We use XSPEC v12.12.0 for the following spectral studies \citep{1996ASPC..101...17A}. We have tried several continuum models, and found that the model {$\tt constant\ast \tt TBabs\ast thcomp\ast diskbb$} is the best-fitting model. In this model, the multiplicative {\tt constant} is used to quantify the cross-calibration uncertainties between the three instruments of \hxmt. 
%
% We have tried different fitting models such as {\tt bbody, powerlaw, cutoffpl, thcomp, simpl} etc., and found good ${\chi} ^{2}$ results for {$\tt TBabs\ast thcomp\ast diskbb$} model. 
%
The component {\tt thcomp} is a novel thermal Comptonization convolution model to describe the broadband X-ray power-law continuum, which is presented as a replacement for the {\tt nthcomp} model \citep{1996MNRAS.283..193Z,2020MNRAS.492.5234Z,2021MNRAS.506.2020D,2022MNRAS.512.4541W}. To use the {\tt thcomp} convolution model, we extend the energy grid from 0.1 keV to 500 keV with the {\tt XSPEC} command {\tt ENERGIES} to treat with the limited response matrices at high or low energies. Because the distance of \target\ is not precisely measured, the hydrogen column density is still not well determined ($N_{\rm H}=(5-12)\times10^{22}~{\rm cm}^{-2}$) \citep{1998ApJ...494..747T,2018ApJ...867...86P}. Given that the column density of the Galactic absorption is not expected to change on short timescales, we fit the five spectra simultaneously with $N_{\rm H}$ linked.

\iffalse The ObsIDs we used in joint fitting are listed in Table~\ref{log} with the mHz QRM phenomenon. The useful energy band for NICER we choose is 1.0--10.0 keV and the 1.0--3.0 keV data includes 1\% systematic errors. As Figure~\ref{NICER_HXMT} shows, the model {$\tt TBabs\ast thcomp\ast diskbb$} satisfies a good fitting result ($\chi^2_{\rm red}=1.14$) and we determine a moderate value $N_{\rm H}=7.67\times10^{22}~{\rm cm}^{-2}$ consistent with previous studies \citep{2001MNRAS.322..309T,2020MNRAS.497.1197B}.\fi
%
The best-fitting spectral parameters are listed in Table~\ref{table_spec}. The errors are given at 1$\sigma$ level. The best-fitting models are shown in Figure~\ref{spec}, along with the residual of each spectrum. From our fits, we get a moderate value of $N_{\rm H}=8.24^{+0.03}_{-0.07}\times10^{22}~{\rm cm}^{-2}$, which is consistent with previous results \citep{2001MNRAS.322..309T,2020MNRAS.497.1197B}.

We first compare the spectra of the period with the type-C QPO and that with the mHz QRM (epochs 1 and 2). From epoch 1 to epoch 2, the disk temperature increases from $1.02^{+0.10}_{-0.10}$ keV to $1.16^{+0.02}_{-0.02}$ keV, accompanied by an increase in {\tt diskbb} normalization. In addition to the disk variation, the electron temperature $kT_{\rm e}$ increases slightly from $20.7^{+2.8}_{-2.0}$ keV to $29.5^{+1.4}_{-1.3}$ keV. Meanwhile, a smaller optical depth is seen in epoch 2 ($\tau_{\rm hot}=1.84^{+0.07}_{-0.07}$) than that in epoch 1 ($\tau_{\rm hot}=2.93^{+0.27}_{-0.35}$).

We then compare the spectra of the period with and without the mHz QRM (epochs 2, 3, 4 and 5). The disk temperature is lower when the QRM appears, while the {\tt diskbb} normalization is slightly higher. The optical depth of the Comptonization region is slightly higher when the QRM appears.

Apart from the differences discussed above, $f_{\rm sc}$ is constrained at its upper limit for epochs 2,3,4 and 5, but its value is slightly smaller in epoch 1.

In order to constrain the minimum set of spectral parameters that dominate the spectral changes, we conduct a simultaneous fit for two sets of spectra: the spectra of the epochs with the type-C QPO and the QRM (epochs 1 and 2), and the spectra of the epochs with and without the QRM (epochs 2 and 3). As \citet{2013ApJ...775...28S} did for type-B QPO, we first tie all parameters together, which leads to a very high $\chi^2/d.o.f.$. We then untie the parameters of {\tt diskbb} and {\tt thcomp} one by one, and check the change of $\chi^2/d.o.f.$ (see Table~\ref{link_spec}). %{\color{red}\textbf{We first untie the parameters of {\tt diskbb}, and then free the parameters of {\tt thcomp} as the first two columns shows in Table~\ref{link_spec}. Afterwards, we change the order in which the parameters are freed, and untie {\tt thcomp}, {\tt diskbb} in sequence as the last three columns shows.}}
%
%We have tried allowing the {\tt diskbb} parameters first or allow the {\tt thcomp} parameters first. 
%
We find that, for both sets of spectra, untying the parameters of the disk component significantly improves the fits. The two parameters of {\tt diskbb}, $T_{\rm in}$ and $norm$, are highly correlated. Therefore, it is hard to distinguish which parameter leads to the changes. Freeing $T_{\rm in}$ only or freeing $norm$ only results in a similar $\chi^2/d.o.f.$. For both sets of spectra, the optical depth of the corona is also needed to be free to obtain an acceptable fit. 
Allowing $kT_{\rm e}$ and $f_{\rm sc}$ to vary only marginally improves the fit for the set epochs 1 and 2, and does not improve the fit any more for the set epochs 2 and 3.  
We have also tried to free the {\tt thcomp} parameters first. However, freeing all the parameters of {\tt thcomp} still leads to an unacceptable fit with  $\chi^2/d.o.f. \gtrsim 2$.

Moreover, due to the relatively low flux of the source at the time of observation, we cannot perform phase-resolved spectral study or time-resolved spectral study for \target\ as done for GRS 1915+105 \citep{2012_neilsen_radiation,2022_Rawat_time}. Nevertheless, we performed a simple flux-resolved spectra comparison for the mHz QRM. We extracted spectra from data above the average count rate and below the average count rate, respectively. We fit the low-flux and high-flux spectra jointly with the model {$\tt constant\ast \tt TBabs\ast thcomp\ast diskbb$}. The results show a slight difference in normalization, while the other parameters remain consistent within $1\sigma$ error.

\iffalse
\begin{figure}
\epsscale{1.0}
\centering
	\includegraphics[width=0.48\textwidth]{spec_NICER_HXMT.pdf}
    \caption{Joint fitting of NICER and \hxmt\ observations of \target\ taken in 2021. The adopted model is {$\tt TBabs\ast thcomp\ast diskbb$}.}
    \label{NICER_HXMT}
\end{figure}
\fi

\begin{figure}
\epsscale{1.0}
\centering
	\includegraphics[width=0.48\textwidth]{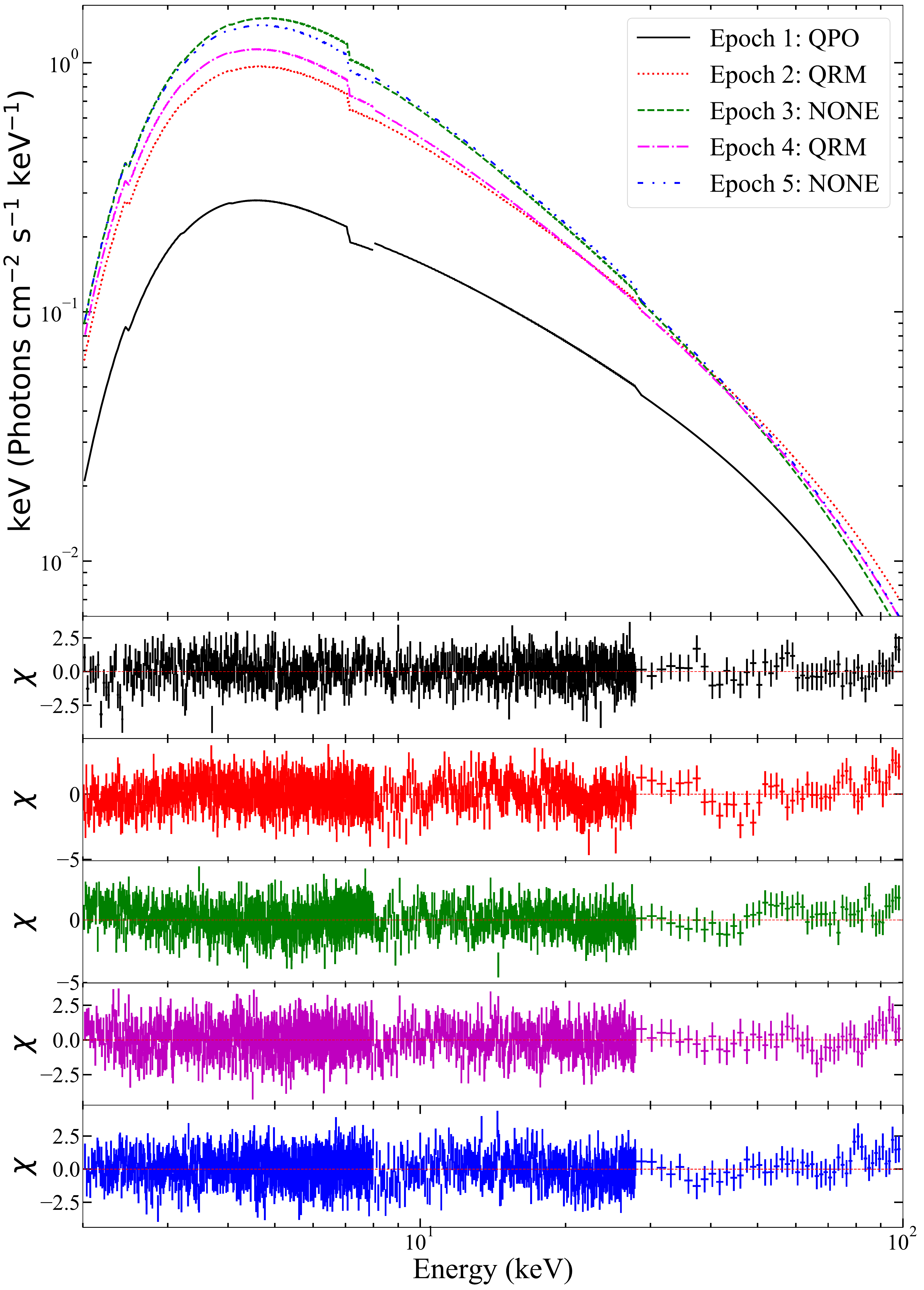}
    \caption{The best-fitting models and the corresponding residuals to the five spectra with model {$\tt constant\ast TBabs\ast thcomp\ast diskbb$}.}
    \label{spec}
\end{figure}

\begin{deluxetable*}{lcccccc}
\tablecaption{Spectral parameters of all epochs using the model {$\tt constant\ast \tt TBabs\ast \tt thcomp\ast \tt diskbb$}. $N_{\rm H}$ is the X-ray absorption column density. $T_{\rm in}$ is the temperature at inner disk radius. Norm is the normalization parameter of $\tt diskbb$. $\tau_{\rm hot}$ is the Thomson optical depth. $kT_{\rm e}$ is the electron temperature. $f_{\rm sc}$ is the scattering fraction. $R_{\rm in}$ is the apparent disk radius, defined as $R_{\rm in}=\sqrt{norm/{\rm cos}\theta}\times D_{10}$, where $D_{10}$ is the source distance in units of 10 kpc and $\theta$ is the inclination angle of the accretion disk. Here we adopt $D_{10}=1$ and $\theta=64\degree$ \citep{King_2014_wind}. The letter F indicates that the parameter was fixed. The letter P indicates that the error of the parameter pegged at the upper boundary. \label{table_spec}}
\tablewidth{700pt}
\tabletypesize{\scriptsize}
\tablehead{
\colhead{Component} & \colhead{Parameter} & \colhead{Epoch 1} & \colhead{Epoch 2} & \colhead{Epoch 3} & \colhead{Epoch 4} & \colhead{Epoch 5} 
}

\startdata
& constant (LE)  & 1.00 (F) & 1.00 (F) & 1.00 (F) & 1.00 (F) & 1.00 (F)\\
& constant (ME)  & 1.07 & 1.00 & 0.91 & 0.97 & 1.05\\
& constant (HE)  & 1.02 & 0.94 & 0.87 & 0.95 & 0.98\\
\hline
TBabs	&	$ N_{\rm H} (\times 10^{22}~\rm cm^{-2})$	&	\multicolumn{5}{c}{$8.24^{+0.03}_{-0.07}$ (link)}\\
\hline
diskbb	&	$ T_{\rm in} (\rm keV)$	&	$1.02^{+0.10}_{-0.10}$	& $1.16^{+0.02}_{-0.02}$  & $1.41^{+0.03}_{-0.03}$ & 	$1.14^{+0.03}_{-0.03}$ & $1.31^{+0.02}_{-0.03}$\\
	&	norm	&	$214^{+110}_{-62}$	& $431^{+35}_{-44}$  & ${312}^{+28}_{-24}$ &  ${547}^{+45}_{-57}$ & $394^{+34}_{-30}$\\
	&   $R_{\rm in} (R_{\rm g}$)    &   $1.21^{+0.21}_{00.18}$ & $1.68^{+0.03}_{-0.03}$ & $1.45^{+0.03}_{-0.03}$ & $1.93^{+0.08}_{-0.07}$ & $1.69^{+0.04}_{-0.04}$\\
	\iffalse
	&   norm($R_{\rm g}$\footnote{$f_{\rm col}=1.7$})    &   $3.52^{+0.62}_{-0.54}$ & $4.85^{+0.09}_{-0.10}$ & $4.19^{+0.11}_{-0.11}$ & $5.60^{+0.24}_{-0.22}$ & $4.89^{+0.13}_{-0.13}$\\
	\fi
\hline
thcomp	& $\tau_{\rm hot}$	&	$2.93^{+0.27}_{-0.35}$	&	$1.84^{+0.07}_{-0.07}$  & $1.67^{+0.12}_{-0.07}$ &  $1.80^{+0.15}_{-0.13}$	& $1.51^{+0.09}_{-0.05}$ \\
 & $kT_{\rm e}$	& 	$20.7^{+2.8}_{-2.0}$	& $29.5^{+1.4}_{-1.3}$  & $28.1^{+2.8}_{-2.5}$ & $28.2^{+3.6}_{-2.6}$ & $31.9^{+2.4}_{-2.2}$\\
 &  $f_{\rm sc}$	&	$0.76^{+0.09}_{-0.09}$	&	$1.00^{+\rm P}_{-0.01}$  & $1.00^{+\rm P}_{-0.01}$ & $1.0^{+\rm P}_{-0.03}$  & $1.0^{+\rm P}_{-0.01}$\\
\hline
${\chi} ^{2}/d.o.f$ & & 572/671 & 1045/1097  & 1054/1085 & 887/1019 & 984/1046\\
\enddata

\end{deluxetable*}

\iffalse
\begin{figure}
\epsscale{1.0}
\centering
	\includegraphics[width=0.48\textwidth]{phase_parameter.pdf}
    \caption{The evolution of spectral parameters vs. epoch. The parameters are, respectively, disk temperature $T_{\rm in}$, disk normalization norm, Thomson optical depth $\tau$, electron temperature $E_{\rm cut}$, the scattering fraction $f_{\rm sc}$. The error bars represents 1$\sigma$ confidence level.}
    \label{evolution_parameter}
\end{figure}
\fi

\begin{deluxetable*}{ccccccc}
\tablecaption{Simultaneous spectral fits using the model {$\tt constant\ast \tt TBabs\ast \tt thcomp\ast \tt diskbb$}. We first link all parameters together, and then untie one parameter at one time to check which parameter is critical to illustrate the spectral differences. see the text for details. \label{link_spec}}
\tablewidth{700pt}
\tabletypesize{\scriptsize}
\tablehead{
\colhead{} & \colhead{} & \colhead{} & \colhead{$\chi^2/d.o.f.$} & \colhead{} & \colhead{} & \colhead{}  
}

\startdata
Parameter & All tied  & $ T_{\rm in} (\rm keV)$ & norm & $\tau_{\rm hot}$ & $kT_{\rm e} (\rm keV)$ & $f_{\rm sc}$\\
\hline
Epoch 1,2 & 30620/1759 & 1825/1758 & 1818/1757 & 1626/1756 & 1624/1755 & 1619/1754 \\            
Epcoh 2,3 & 11200/2173 & 2623/2172 & 2622/2171 & 2102/2170 & 2102/2169 & 2102/2168 \\
\hline
Parameter & All tied   & norm  & $ T_{\rm in} (\rm keV)$ & $\tau_{\rm hot}$ &  $kT_{\rm e} (\rm keV)$ & $f_{\rm sc}$\\
\hline
Epoch 1,2 & 30620/1759 & 1827/1758 & 1818/1757 & 1626/1756 & 1624/1755 & 1619/1754 \\            
Epcoh 2,3 & 11200/2173 & 2676/2172 & 2622/2171 & 2102/2170 & 2102/2169 & 2102/2168 \\
\hline
Parameter & All tied     & $\tau_{\rm hot}$ &  $kT_{\rm e} (\rm keV)$ & $f_{\rm sc}$ & $ T_{\rm in} (\rm keV)$ & norm \\
\hline
Epoch 1,2 & 30620/1759 & 4679/1758 & 4017/1757 & 3314/1756 & 1621/1755 & 1619/1754 \\            
Epcoh 2,3 & 11200/2173 & 6742/2172 & 5143/2171 & 5143/2170 & 2118/2169 & 2102/2168 \\
\hline
Parameter & All tied     &  $kT_{\rm e} (\rm keV)$ & $\tau_{\rm hot}$ & $f_{\rm sc}$ & $ T_{\rm in} (\rm keV)$ & norm\\
\hline
Epoch 1,2 & 30620/1759 & 5516/1758 & 4291/1757 & 4291/1756 & 1624/1755 & 1619/1754 \\            
Epcoh 2,3 & 11200/2173 & 11122/2172 & 4726/2171 & 4726/2170 & 2118/2169 & 2102/2168 \\
\hline
Parameter & All tied     &   $f_{\rm sc}$ & $kT_{\rm e} (\rm keV)$ & $\tau_{\rm hot}$ & $ T_{\rm in} (\rm keV)$ & norm \\
\hline
Epoch 1,2 & 30620/1759 & 3483/1758 & 3315/1757 & 3315/1756 & 1621/1755 & 1619/1754 \\            
Epoch 2,3 & 11200/2173 & 6142/2172 & 5570/2171 & 4619/2170 & 2118/2169 & 2102/2168 \\
\enddata

\end{deluxetable*}

\iffalse
\begin{figure}
\epsscale{1.0}
\centering
	\includegraphics[width=0.48\textwidth]{model.pdf}
    \caption{Cartoon picture showing the evolution of the accretion disk and corona. In epoch 1, an ellipsoid-like corona sandwiches the disk and performs Lense-Thirring precession to produce the type-C QPO. In epoch 2 and epoch 4, with the increasing luminosity and expansion of corona, the ellipsoid-like corona evolves to a spherically symmetric corona. Almost all seed photons from accretion disk are Comptonised. An unknown instability in corona-disk coupling region leads to the regular flux modulation. In epoch 3 and epoch 5, the conditions for this instability are broken because of increasing accretion rate and inner temperature of accretion disk.}
    \label{model_cartoon}
\end{figure}
\fi

\section{DISCUSSION} 

In section 3, we have carried out a detailed timing and spectral analysis of the five epochs with type-C QPO and the secular mHz QRM in 4U1630-47, as observed by \hxmt\ during the 2021 outburst. In this section, we will discuss the possible mechanisms behind the mHz QRM based on above results.

%We find that the appearance of the mHz QRM depends on the source luminosity. After comparing the spectra when the mHz QRM appears and disappears, we find obtain a few interesting results occurring when specific transition happens. 

\subsection{Appearance of the mHz QRM}

In terms of 2021 outburst, the mHz QRM appears shortly after the disappearance of type-C QPO within one day (from MJD 59482.626 to MJD 59483.785), along with a drop of total fractional rms (see Figure~\ref{HXMT}). In the bottom panel of Figure~\ref{HXMT}, the fractional rms remains stable before MJD 59483 when the type-C QPO is prominent in PDS. After MJD 59483, the rms amplitude decreases from above 25\% to below 20\% when the mHz QRM appears. The drop of fractional rms during intermediate state is usually accompanied by the transition from hard intermediate state to soft intermediate state.

A similar mHz QRM was also observed in the 1998 outburst of \target\ with \RXTE\ \citep{2001MNRAS.322..309T}. However, the light curves of the mHz QRM seen in the 1998 and 2021 outbursts are quite different. For the mHz QRM in the 1998 outburst, the light curve is characterized by the presence of quasi-regular dips with a period of $\sim$10--20 s, instead of the flare-like features shown in Figure \ref{lc_pds}.
%
%Unlike in the 1998 outburst, the type-C QPO and mHz QRM are detected from different PDS in the 2021 outburst. 
%
In particular, the PDS of the observation of the 1998 outburst show three extra prominent QPO features at $\sim$4.7 Hz, $\sim$7.0 Hz and $\sim$13 Hz, in addition to the peak of the mHz QRM \citep{2000_ApJ_4U1630,2001MNRAS.322..309T}. Further, \citet{2001MNRAS.322..309T} investigated the flux-dependent properties of this observation. They divided the observation into two different flux levels and found that the $\sim$13 Hz QPO is prominent at low fluxes, and disappears at high fluxes. The $\sim$4.7 Hz QPO was observed at low fluxes, and shifted to $\sim$7 Hz at high fluxes. However, in the 2021 outburst, the type-C QPO is not simultaneously detected with the mHz QRM. We have also made a comparison of the PDS between the low-flux period and the high-flux period for our observations with the mHz QRM. No significant QPO signal was found in the 1--20 Hz range. To check the possibility of the presence of a low-frequency QPO, we add an extra Lorentzian component to the PDS with the frequency fixed at any potential peak in the frequency range 1--10 Hz, and get a very low rms value ($<1\%$). Therefore, we can rule out the presence of a significant low-frequency QPO when the mHz QRM appears.

Meanwhile, the appearance of mHz QRM is related to the accretion rate based on our results (see Figure~\ref{dyps}), corresponding to $L_{2-100 \rm \ keV} \sim 0.16 \ L_{\rm Edd}$ (assuming a distance of $D=10 \ {\rm kpc}$).
%
%which has been previously found in GRS 1915+105, GRO J1655-40, IGR J17091--3624 \citep{1997ApJ...482..993M,1999ApJ...522..397R,2017MNRAS.468.4748C}. 
We have also checked the \RXTE\ observations of the 1998 outburst and confirmed that the mHz QRM is also observed in a special flux plateau.
\citet{2001MNRAS.322..309T} suggested that the accretion-rate dependent flares might be common for black hole binaries emitting at a certain luminosity level. For \target, the appearance of the mHz QRM was observed at a very similar luminosity and position in HID compared to the 1998 outburst: the mHz QRM appears at a flux of $1.37\times10^{-8} \rm \ erg \ s^{-1} \ cm^{-2}$ in 3--20 keV for 2021 outburst and at $1.40\times10^{-8} \rm \ erg \ s^{-1} \ cm^{-2}$ for 1998 outburst. This behaviour can be explained by the local radiation dominated disk instability model, considering the occurrence of limit cycle behaviour under certain narrow accretion-rate range \citep{2008bhad.book.....K}.
Phenomenologically, the quasi-regular flare can be interpreted as the rapid removal and replenishment of matter forming the inner region of the accretion disk, as observed in GRS 1915+105 \citep{1997_ApJ_Belloni}. This process finishes a cycle on the viscous/thermal timescale.
\iffalse To verify this model, we compare the spectra when counts rate is greater than average counts rate and when counts rate is less than average counts rate and get  the inner radius of accretion disk is $R_{\rm in}^{\rm high}=27^{+5}_{-3} R_g$, $R_{\rm in}^{\rm low}=17\pm 2 R_g$.\fi 
 
%

Although the light curves of the mHz QRM seen in the 1998 and 2021 outbursts show different variations, the energy-dependent fractional rms and lag of the mHz QRM are similar (Zhao et al. in preparation). Considering that the mHz QRM appears in a similar flux range, we suggest that the mHz QRM seen in the two outbursts may have a similar physical origin.

Apart from the comparison with the 1998 outburst, we also compare this mHz QRM with the $\sim$11 mHz QPO detected in H 1743--322 \citep{2012ApJ...754L..23A,2019MNRAS.482..550C}. \citet{2012ApJ...754L..23A} reported the detection of a $\sim$11 mHz QPO in H 1743--322 at the beginning of the 2010 and 2011 outbursts. The mHz QPO was found at a similar spectral hardness value and intensity level, suggesting an accretion-rate dependence. We can see clear quasi-regular modulations from the light curves, similar to what we find in \target. However, the $\sim$11 mHz QPO is much narrower ($Q \sim 10$) than the QRM observed in \target. In addition, the fractional rms amplitude of the $\sim$11 mHz QPO is significantly lower ($3.1\%\ \pm\ 0.4\%$). Meanwhile, its energy-dependent properties also seems to be different from that found in \target\ \citep{2012ApJ...754L..23A,2019MNRAS.482..550C}. Therefore, it is not clear whether the two mHz signals have a similar physical origin or not.

\subsection{Properties of the mHz QRM}
 
Thanks for the large effective area and broad energy range of \hxmt, we are able to investigate the energy dependence of the QRM properties to higher energy band. From Figure~\ref{E-rms-f}, we can see that both the fractional rms and characteristic frequency of the QRM increase with energy up to 60--100 keV. This positive correlation is more significant in fractional rms than in frequency, inferred from the results of linear function fitting. 
\iffalse In general, the harder energy photon can be associated with a smaller radius in the inner region of accretion flow. Then the viscous/thermal timescale for this region will be smaller than outer region. In other words, the emptying and refilling of material occurring here will repeat more frequently than softer photon.\fi 
The energy-dependent fractional rms provides important evidence to constrain the physical origin of the mHz QRM. The increasing fractional rms up to 60--100 keV implies that the mHz QRM phenomenon is more pronounced for no-thermal components, the so-called corona/hot flow region. 
%
%{\color{red}\textbf{Therefore, it seems that this phenomenon comes directly from the high energy radiation region, yet still cannot fully exclude the possibility of the accretion-disk instability model.}} 
%
It is worth noticing that some sub-Eddington sources, such as IGR J17091--3624, also show this exotic variability while other Eddington-limited sources, such as GX 17+2 and V404 Cyg, do not \citep{2017MNRAS.468.4748C}.  \citet{2016MNRAS.462..960S} concluded that the disk size or minimum stabilizing large-scale magnetic field may be the unifying factor behind the objects that display GRS 1915-like variability. In conclusion, the results we obtained can well constrain the origin of the mHz QRM to the hot corona region.    

The spectral evolution between different epochs provide more information on the physical origin of the transitions. 
%
%In the 2021 outburst, we detected significant emission above 30 keV by \hxmt, which is similar to the 2010 outburst rather than the 2006 and 2008 outburst. As \citet{2015MNRAS.450.3840C} found, the 2010 outburst extends at high energies without any detectable cut-off until 150--200 keV, whereas the two previous outbursts that occurred in 2006 and 2008 are not detected at all above 30 keV. For the 2021 outburst, we not only detect significant emission above 30 keV, but also confirm a consistent electron temperature $E_{\rm cut} \sim 20-30 \ \rm keV$ at all 5 epochs as shown in Table~\ref{table_spec} due to the capability of detecting emission above 30 keV for \hxmt. 
%
We find two transitions: from the epoch with the type-C QPO to the appearance of the mHz QRM, and from the epoch with the mHz QRM to the disappearance of the mHz QRM. 
%
%One of the transitions takes place between epoch 1 and epoch 2, another transition takes place between epoch 2 and epoch 3, epoch 4 and epoch 5. 
%
The transition from the type-C QPO epoch to the mHz QPO epoch appears to be accompanied by an increase in inner disk radius. By comparing the optical depth of corona and scattering factor $f_{\rm sc}$ between epochs 1 and 2, we find that only $\sim$76\% of the seed photons are scattered in epoch 1, whereas nearly all photons from accretion disk are scattered in the Comptonization region in epoch 2. Meanwhile, the optical depth of the corona decreases from $\tau_{\rm hot}=2.93^{+0.27}_{-0.35}$ to $\tau_{\rm hot}=1.84^{+0.07}_{-0.07}$. Compared to the period without the mHz QRM, we find a colder inner disk and a larger disk radius when the mHz QRM appears in epoch 2 and 4. In addition to changes in {\tt diskbb} parameters, we also find that, when the mHz QRM appears, the optical depth $\tau_{\rm hot}$ is slightly higher. 
To summarize, the disk parameters and the optical depth have a significant effect on the goodness of the fits.

\iffalse
We speculate that the decrease of optical depth means the expansion of corona to cause a decrease of electron density \citep{2020MNRAS.492.5234Z}. The expansion of corona will also increase coupling area of disk and corona, thus lead to a fully scattered Comptonization spectrum.This relation between the Comptonization component and the mHz QRM implicates that the change of corona optical depth, inner temperature of accretion disk may be connected with the appearance of the special mHz QRM phenomenon. We speculate that the joint region where inner part of accretion disk interacts with corona could be the place this flux modulation happens. 
\fi

Meanwhile, the more or less zero lags below 10 keV and negative lags above 10 keV put strict restrictions on the origin of this mHz QRM. 
\iffalse But we also notice that the negative time lag as low as -3s may constrain tightly the origin of this mHz QRM despite only for the highest energy band 60--100 keV with very  Figure~\ref{timelag_evolution}.\fi 
The magnitude of the time delays on timescales of seconds implies that they can not be produced by light travel time effects or Comptonization \citep{2011MNRAS.414L..60U}. Considering the fast velocity of jet-like ejection ($v \sim0.1c$), we cannot attribute the change of time delay to the change of the size of ejection. If it is the case, its size will be larger than $10,000~R_{\rm g}$. 
%
%Therefore, in order to explain the soft time delay above 10 keV, we need a new scenario to explain the soft time lag.
\citet{2016_MNRAS_Mir} used the \RXTE/PCA archival data of GRS 1915+015 during the "$\rho$" class, and produced the energy-dependent rms and time lag for the "heartbeat" modulation. They found that the time lags turn to a large soft lag at high energies, similar to what we found in \target\ for the mHz QRM.
However, the fractional rms turns around at $\sim$10 keV, and then decreases to a lower value with energy. This is opposite to the positive correlation we observed in \target. 
\citet{2016_MNRAS_Mir} proposed that there is a delayed response of the inner disk radius to the accretion rate. In this model, the fluctuating accretion rate in the outer disk drives the oscillations of the inner radius after a time delay up to several seconds, $r_{in}(t)\propto \dot{m}^\beta (t-\tau_d)$, while the power-law component responds immediately. They showed that, in such a scenario, a pure sinusoidal oscillation of the accretion rate can explain the shape and magnitude of the energy dependent rms and time lag of the "heartbeat" modulation seen in GRS 1915+015.
%
%
%with several parameters, such as the power-law to disk flux ratio parameter $N$, parameter $\beta$ which relates the dependence of the inner radius with the accretion rate, power-law index $p$. 
However, in the case of \target, by broad band spectral analysis, we find that the spectra can be well-fitted by a single thermal Comptonization component ($f_{\rm sc} = 1$) which fully scatters the seed photon from a multi-black body accretion disk. This means that the radiation from 1-100 keV is almost entirely from the Compton scattering of the corona. Meanwhile, the energy-dependent rms shown in Figure~\ref{E-rms-f} cannot be fitted under the model that \citet{2016_MNRAS_Mir} presented.
%
%Thus, the simple model can explain the energy-dependent time lag but can not predict the energy-dependent rms for \target.
%
A further investigation on the model is needed in the future to explain the results shown in our work. 
From epcoh 1 to epoch 2, the decreasing optical depth suggests the expansion of the corona, leading to a decrease in electron density \citep{2020MNRAS.492.5234Z}. 
%
%because of optical depth $\tau=n_{\rm e}\sigma_{\rm T}R$ 
%
The expansion of the corona will also increase the coupled area between the disk and corona, thus leading to a fully scattered Comptonization spectrum and an increase in scattering fraction. %With the increasing luminosity and expansion of corona, the ellipsoid-like corona evolves to a spherically symmetric corona, similar to the morphology of the corona in \citet{2021MNRAS.503.5522K}. 
We speculate that the joint region where the inner part of the accretion disk interacts with corona could be the place this flux modulation happens. 
An unknown instability with strict accretion rate dependence in the corona leads to the quasi-regular flux modulation . If the luminosity is slightly off the narrow luminosity range, the quasi-regular flux modulation will not appear to be accompanied with increasing disk temperature, which may reconstruct the coupling region between corona and accretion disk.
%
%{\color{red}\textbf{However, we should also note that the distinct timing properties for the 1998 outburst and this 2021 outburst, especially for the appearance of LFQPOs. Then the geometry of the accretion flow may be very different for the two outbursts. As a result, the establishment of more refined model needs more QRM samples in the future.}} 

\iffalse
According to discussion above, we present a toy model as Figure~\ref{model_cartoon}. When only the type-C QPO appears, the "sandwich" corona lies over accretion disk (the inner radius of accretion disk is almost at ISCO). Then with increasing accretion rate, the corona disappears (may condensate to accretion disk or be blown up). Meanwhile, the radiation pressure blows up the material provided by periodic accretion rate change. This material maybe collimated by magnetic field to a hot ejection. The hot ejection is highly ionised, hot and acts as non-thermal emission contributor.
\fi

\section{CONCLUSION} \label{sec:highlight}

In this work, we present a detailed energy-dependent timing analysis and spectral comparison for the mHz QRM in \target. The main results are summarised as follows:

(1) The fractional rms of the mHz QRM increases with photon energy from 1 to 100 keV. The characteristic frequency of the mHz QRM increases marginally with energy. The phase/time lag of the mHz QRM is more or less zero below 10 keV and becomes negative above 10 keV. The absolute lag increases with energy from 10 keV up to 60--100 keV.

(2) The behaviour of the mHz QRM significantly depends on the accretion rate. 
The corresponding flux in 3--20 keV is about $1.37\times 10^{-8} \ \rm erg \ s^{-1} \  cm^{-2}$, which is very close to that observed in the 1998 outburst. 
%This is the second time this phenomenon has been detected.

(3) We measure the spectral differences among the periods with the type-C QPO, with the mHz QRM and without the mHz QRM. Compared with the type-C QPO spectra, the inner disk temperature increases, while the optical depth of the corona decreases when the mHz QRM appears. Meanwhile, the scattering fraction increases marginally from type-C epoch to mHz QRM epoch. Compared with the period without the mHz QRM, the optical depth of the corona is higher, while the inner disk temperature is slightly lower when the mHz QRM appears.

This mHz QRM phenomenon could be interpreted by an unknown instability in corona that causes the flux modulation. Meanwhile, the change of corona optical depth/size and inner temperature of disk may be related to the appearance of the mHz QRM. However, a more complex model is needed to explain the energy-dependence rms and time lag. Observations of the mHz QRM with future X-ray telescopes, e.g., the enhanced X-ray Timing and Polarimetry mission (eXTP, \citealt{2019SCPMA..6229502Z}), are required to perform a more detailed time/phase-resolved analysis and give a deeper insight into its physical origin.

%Meanwhile, the analysis of frequency/phase resolved spectra can help us to shed light on the physics mechanism behind this mHz QRM phenomenon.

\acknowledgments
We are grateful for the anonymous referee's helpful comments and suggestions. This work has made use of the data from the \hxmt\ mission, a project funded by China National Space Administration (CNSA) and the Chinese Academy of Sciences (CAS), and data and/or software provided by the High Energy Astrophysics Science Archive Research Center (HEASARC), a service of the Astrophysics Science Division at NASA/GSFC. This work is supported by the National Key RD Program of China (2021YFA0718500) and the National Natural Science Foundation of China (NSFC) under grants U1838201, U1838202, 11733009, 11673023, U1938102, U2038104, U2031205, the CAS Pioneer Hundred Talent Program (grant No. Y8291130K2) and the Scientific and Technological innovation project of IHEP (grant No. Y7515570U1).
 
%We thank the referee for the fruitful comments.

\textit{Software}: XSPEC \citep{1996ASPC..101...17A}, Astropy \citep{2013A&A...558A..33A}, Numpy \citep{2011CSE....13b..22V}, Matplotlib \citep{2007CSE.....9...90H}, Stingray \citep{2019ApJ...881...39H,2019JOSS....4.1393H}.
%%%%%%%%%%%%%%%%%%%%%%%%%%%%%%%%%%%%%%%%%%%%%%%%%%
\section*{Data Availability}
The raw data underlying this article are available at http://hxmten.ihep.ac.cn/.


\vspace{5mm}

\bibliographystyle{aasjournal}
\bibliography{reference}{}



\end{document}